\documentclass[12pt]{article}
\usepackage[utf8]{inputenc}
\usepackage[a4paper, total={6in, 8in}]{geometry}
\usepackage{multirow}
\usepackage{enumitem}
\usepackage{multirow}
\usepackage{float}
\usepackage{colortbl}
\usepackage{caption}
\usepackage{subcaption}
\usepackage{dsfont}
\usepackage{graphicx}
\title{Scale matters: The daily, weekly and monthly volatility and predictability of Bitcoin, Gold, and the S\&P 500}
\usepackage{color,soul}
\author{Nassim Dehouche}
\usepackage{hyperref}
\usepackage{amssymb}
\usepackage{multirow}
\date{}
\begin{document}

\maketitle
\section*{Abstract}
A reputation of high volatility accompanies the emergence of Bitcoin as a financial asset. This paper intends to nuance this reputation and clarify our understanding of Bitcoin's volatility. Indeed, two distinct and non-redundant understanding of volatility as deviation from consistency exist for a time-series: (1) exhibiting high standard deviation, and (2) appearing highly irregular or unpredictable. Using daily, weekly, and monthly closing prices and log-returns data going from September 2014 to January 2021, we find that Bitcoin is a prime example of an asset for which the two conceptions of volatility diverge. We show that, historically, Bitcoin allies both high volatility (high Standard Deviation) and high predictability (low Approximate Entropy), relative to Gold and S\&P 500.  

Moreover, using tools from Extreme Value Theory, we analyze the convergence of moments, and the mean excess functions of both the closing prices and the log-returns of the three assets. We find that the closing price of Bitcoin is consistent with a generalized Pareto distribution, when the closing prices of the two other assets (Gold and S\&P 500) present thin-tailed distributions. However, returns for all three assets are heavy tailed and second moments (variance, standard deviation) non-convergent. In the case of Bitcoin, lower sampling frequencies (monthly vs weekly, weekly vs daily) drastically reduce the Kurtosis of log-returns and increase the convergence of empirical moments to their true value. The opposite effect is observed for Gold and S\&P 500; tails become progressively heavier and sample standard deviations higher and less convergent, the lower the sampling frequency. Thus, the weekly log-returns of Bitcoin present both a lower Kurtosis, Approximate Entropy, Coefficient of Variation (Standard Deviation to Mean ratio), and faster convergence of moments than Gold and S\&P 500. These properties suggest that Bitcoin's volatility is essentially an intra-day and intra-week phenomenon that is strongly attenuated on a weekly time-scale, and make it an attractive store of value to investors and speculators, but its high standard deviation excludes its use a currency.

\section{Introduction and related work}
 Owing to its ability to act as an inflation-resistant store of value  and a decentralized medium of exchange, Bitcoin has recently reached a market capitalization of \$1 trillion \cite{cnbc} and cemented its emergence as a distinct financial asset \cite{krueckeberg}. This exponential increase in value has however come with a reputation of persistent volatility reflected in indicators based on the standard deviation of Bitcoin's daily price compared to traditional financial assets (e.g. Gold, S\&P 500).

For both Finance academics \cite{vol1, vol2} and practitioners\cite{vol3}, common measures of price volatility are based on the standard deviation (e.g. in GARCH models \cite{dyhrberg}) of the price or returns of an asset relative to its simple moving average.
 
 Due to the extreme in its price, Bitcoin is heterologically used as digital asset rather than a currency. Using a GARCH model, \cite{dyhrberg} found the cryptocurrency to possess hedging capabilities that sets the cryptocurrency in between Gold and the US dollar. Katsiampa \cite{katsiampa} compared different GARCH models for the estimation of the volatility of Bitcoin's daily price.  Bariviera et al. \cite{bariviera} studied the intra-day (sampling frequency: 5 hours) volatility of Bitcoin, relative to Gold and the Euro and the volatility of Bitcoin was found to be decreasing over time, and its long range memory is not related to market liquidity. Moreover, this behavior across different time-scales of 5 to 12 hours was found to be essentially similar, but no longer time-scales were considered. 
 These properties limit its use as currency and Bitcoin is typically regarded as a (digital) commodity. It is classified as such by the Commodity Futures Trading Commission (CFTC) \cite{cftc} and Shahzad et al. \cite{shahzad} find that it shares "safe-haven" properties with gold, and other commodities against fluctuations in global stock market indices. The fact that the extreme movement upward in the price of the cryptocurrency (e.g. its closing price has increased by 14,414\% between 15-09-2014 and 21-02-2021) exclude its use as such, a currency. Our intent is to distinguish these variations from the common parlance and finance notions of \textit{volatility}. 
 
 In this paper, we argue that Bitcoin's price volatility measured with standard deviation-based indicators  has received disproportionate attention when it is only a partial reflection of volatility. Indeed standard deviation measures a time series' dispersion from its average, regardless of the predictability of its deviations. Moreover, the variations in Bitcoin's price are typically studied intra-day and intra-week (Cf. the "Price Volatility" category of literature in the exhaustive review of Bitcoin research of Corbet et al. \cite{corbet}) and its properties for these sampling frequencies are assumed to be be inherited by weekly and monthly price variations. Using Approximate Entropy (ApEn) \cite{pincus91} we propose a complementary, non-redundant perspective on Bitcoin's volatility, relative to two other financial assets, Gold and the S\&P 500 Index. Moreover, we study both closing prices and returns for different sampling frequencies (daily, weekly, and monthly) and find drastic changes in the statistical properties and reduction in the volatility of Bitcoin for weekly data, when more traditional assets show the opposite dynamic for longer time-preferences.  
 
  The remainder of this paper is organized as follows.  Section \ref{toy} discusses some common limitations of standard deviation as measure of volatility, Section \ref{datamethods} describes the data and methods used in this study, and Section \ref{desc} presents descriptive statistics for the closing price and returns of the three considered assets. The main contributions of this paper are presented in Section \ref{results} and consist in an analysis of the tail properties and Approximate Entropy of the aforementioned random variables. Finally, Section \ref{conclusion} concludes this paper with a discussion of the implications and limitations of these findings and their potential value to speculators and traders considering Bitcoin as an investment.

\section{Some limitations of standard deviation as a measure of volatility}
\label{toy}
\subsection{It only measures an extent of variation, not regularity}
  The Cambridge definition of the word \textit{volatile} is “likely to change suddenly and unexpectedly, especially by getting worse”. According to \cite{pincus04}, two distinct and non-redundant forms of deviation from consistency exist for a time-series: (1) exhibiting high standard deviation, and (2) appearing highly irregular or unpredictable, which corresponds to the common understanding of the term. This distinction has important consequences. Standard deviation only measures an extent of deviation from the mean. It does not reflect the suddenness and unexpectedness of these deviation.

Moreover, standard deviation amalgamates positive and negative deviations when, in common parlance, volatility is understood as deviations for the worst. Information theoretic measures, such as Approximate Entropy (ApEn) \cite{pincus91} offer a way to complement these standard deviation by quantifying regularity in a time-series. Additionally, significant increases in the Approximate Entropy of a time-series have been found \cite{pincus04} to foretell major variations in a time-series.\\

Let us consider the following toy example of two time-series, of mean $0$, consisting of two different permutations of the same set of values in $\{-1,0,+1\}$ : 

$$X=(0,+1,0,-1,0,+1,0,-1,0,+1,0,-1)$$
$$Y=(-1,+1,0,0,0,+1,0,-1,0,-1,1,0)$$

$X$ is a perfectly predictable time-series regularly alternating from $-1$ to $0$ to $+1$. $Y$ is a random permutation of $X$ and is, by all means, more volatile. However, the two time series series show the same standard deviation of $0.725$. The regularity of $X$ is, on the other hand, captured by its lower ApEn of $-0.001$, when $Y$ presents an ApEn of $0.471$. Despite its perfect regularity, the standard deviation of $X$ could be made arbitrarily larger than that of $Y$ by replacing $+1$ in the time-series with any large number. One can also very easily generate completely random $\{-1,0,1\}$ time-series of an arbitrarily lower standard deviation than $X$.\\

\subsection{It may be impossible to empirically estimate}
For many assets, the distribution of prices and returns are known to exhibit heavy-tailed behavior \cite{sornette} that is best modeled by a generalized Pareto distribution \cite{coles}, in which empirical moments are dominated by extreme values. This has been shown to be the case for the log-returns of the S\&P 500 index \cite{spfat}, Gold \cite{goldfat}, and Bitcoin \cite{btcfat}.

Consequently, the second moments of the corresponding random variable may not converge and measures of volatility based on sample standard deviation may be uninformative. Because it only quantifies the total “amount” of variation from the average, sample standard deviation is also highly sensitive to extreme variations. The very use of the term “standard” deviation assumes that there is such a thing as a representative deviation, which requires these deviations to be somehow regular. A time-series such as $(0,1,0,-1,0,1,-1,0,10000)$ would exhibit an empirical standard deviation of $3333.33$, but this number can decently not be considered a good estimator of the second moment of the underlying random variable.

\section{Data and Methods}
\label{datamethods}
\subsection{Data}

We use weekly and monthly Bitcoin and gold price data from 17 September 2014 (the earliest listing of Bitcoin on Yahoo Finance) to 16 January 2021 from Yahoo Finance \cite{yahoobtc, yahoogold, yahoogspc}, which represent a total of 6942 daily, 984 weekly, and 231 monthly observations. Log-returns are calculated by taking the natural logarithm of the ratio of two consecutive prices.

Concerning daily price and returns variations, it should be noted that Bitcoin trades 7 days per week, as opposed to Gold and S\&P 500 only trading on weekdays. We consider that there is no price movement during the weekend for the two latter assets, thus somehow artificially reducing their volatility/irregularity relative to Bitcoin, which further reinforces our conclusions.

\subsection{Methods}

\subsubsection{Mean Excess Functions}
For a non-negative random variable $X$, with support in $D(X)$, the excess distribution  over a threshold $a \in D(X)$ is defined \cite{mef1, mef2, mef3} as follows: $$F_a(x)=P(X-a \leq x | X>a), a \in D(X)$$ Intuitively, its complement $1-F_a(x)$ measures the likelihood of $X$ exceeding $a+x$, given that $X$ has exceeded $a$. For instance, if $X$ measures a closing price, $1-F_a(x)$ is the likelihood of the price gaining $x$ more units, given that it has reached $a$ monetary units so far. The Mean Excess ($ME$) function, also known as the Mean Residual Life function, is the expectation of this distribution for random variables of finite expectations and is defined as $$ME(a)=E(X-a|X>a)= \frac{\int_{a}^{+\infty}(x-a) \,dF(x)}{\int_{a}^{+\infty}dF(x)}, a \in D(X)$$ 

Thus, the empirical ME function $ME_n(a)$ of a sample of $n$ observations $X_1, X_2, \dots, X_n$ can be computed by dividing the total amount of excess over a threshold $a \in D(X)$ by the number of observations realizing such excesses, as follow:
$$ME_n(a)=\frac{\sum\limits_{i=1}^{n}(X_i-a)\cdot \mathds{1}_{(X_i>a)}}{\sum
\limits_{i=1}^{n} \mathds{1}_{(X_i>a)}}$$

The excess distribution and mean are the foundations for peaks over threshold (POT) modeling \cite{mef2} which fits distributions to data on excesses and has wide applications notably in risk management, actuarial science, and project management. Moreover, they define three classes of random variables whose mean excess functions exhibits crucially different statistical behaviors:
\begin{itemize}
    \item A decreasing mean excess function is characteristic of thin-tailed random variables with memory.  Gaussian or Poisson random variables possess this property. 
    \item A constant mean excess function is characteristic of \textit{memorylessness} \cite{memoryless}. Exponential random variables and their discrete analogues, Geometric random variables, notoriously exhibit this property. 
       \item An increasing mean excess function is characteristic of scalable heavy-tailed random variables, among which linear ME functions characterize the Generalized Pareto Distribution class \cite{evt1}, whereas convex ME functions are indicative of log-normality \cite{cirillo}. Both classes of random variables can exhibit infinite or slowly converging moments.
       
\end{itemize}

Chaotic perturbations are commonly observed at the extremity of plots of empirical ME functions, as a result of finite sample bias, i.e. the fact that points for very high order statistics in the plot are the result of very few observations. This bias is commonly addressed by discarding points in the plot for very high order statistics \cite{ghosh}.

\subsubsection{Maximum to sum ratios}

 For an order $p \in \{1,2,3,4,\dots\}$, the convergence of the ratio of the maximum to the sum of exponent $p$ is indicative of the existence of the moment of order $p$, and if so of the speed of convergence of the empirical moments of order $p$ to its true value, per the Law of Large numbers. Formally, given a sample of $n$ observations $\{X_1,\dots,X_n\}$ of a positive random variable $X$, let $M(n,p)=Max\{X_1^p, \dots, X_n^p\}$ be the maximum of order $p$ and $S(n,p)=\sum\limits_{i=1}^n X_i^p$, the sum of order $p$. We have the following result \cite{evt1, cirillo, cirilloplos}:

$$E(X^p)< +\infty \Leftrightarrow \lim_{n\to +\infty} \frac{M(n,p)}{S(n,p)}=0$$

Based on the previous equivalence, plots of the Maximum to Sum ratios represent the ratio  
of the maximum to sum of order $p$ as a function of the number of data points for different values of $p$ and indicate a convergence of the moments of order $p$ to a finite value if and only if the ratio converges to zero. The non-convergence of moments, or their very slow convergence (which requires extremely large sample sizes), renders the estimation of statistics such as the Mean, variance/standard deviation, or Kurtosis (respectively, the first, second, and fourth moment) from empirical observations uninformative.

\subsubsection{Approximate entropy}
Approximate Entropy (ApEn), introduced by Pincus \cite{pincus91, pincus92, pincus04}, is a family of information-theoretic statistics quantifying the "extent of randomness"  in a continuous-state processes, given a time-series of $N$ equally-spaced in time observations $u(1),u(2),\dots,u(N)$, and two parameters $m$, a positive integer representing the length of successive observations to be compared, and $r$, a positive real representing a tolerance level. Widely validated and commonly used parameter values for the computation of $ApEn(m,r,N)$ are $m=2$ and $r=20\%$ of the Standard Deviation of the considered time-series \cite{pincus04}. Approximate Entropy assigns a non-negative number $ApEn(m,r,N)$ to the time-series, with larger values corresponding to greater randomness or irregularity and smaller values corresponding to more instances of repetitive patterns of variation. 

Formally, $ApEn(m,r,N)$ is computed using the following algorithm:
\begin{enumerate}
    \item Compute a sequence of real $m$-dimensional vectors $x(1),x(2),\dots x(N-m+1) \in  \mathbb{R}^m$ such that $x(i)=[u(i),u(i+1),\dots,u(i+m-1)]$.
    \item For each $i\in \{1,\dots,N-m+l\}$, compute $C_i^m(r)=$ number of $x(j):j\in \{1,\dots,N-m+l\}$ such that $d(x(i),x(j)) \leq \frac{r}{N-m+1}$, where $d(x(i),x(j))$ is a distance metric between vectors $x(i)$ and $x(j)$ given by: $d(x(i),x(j))=\max\limits_{1 \leq k \leq m}(|u(i+k-1)-u(j+k-1)|)$.
    \item Compute $\Phi^m(r)={(N-m+1)}^{-1}\cdot{\sum\limits_{i=1}^{N-m+1}log \: C_i^m(r)}$.
    \item Compute $ApEn(m,r,N)=\Phi^m(r) - \Phi^{m+1}(r).$  
\end{enumerate}

Thus, $ApEn(m,r,N)$ measures the the logarithmic empirical likelihood that observations that are close (within $r$) for $m$ successive observations remain close (within the same tolerance $r$) on the next incremental
comparison \cite{pincus92, delgado}. 

The presence of repetitive patterns of fluctuation  in a time-series makes it more predictable than a time series in which such patterns are absent. ApEn reflects the likelihood that similar patterns of observations (within an average distance of $r$) will not be followed by additional similar observations. A low ApEn reflect a high frequency of observing previously observed repetitive patterns and thus a more predictable process. An important property of ApEn is that its calculation is model-independent \cite{pincus04}. It is able to quantify the regularity/predictability of time-series data without making any assumptions concerning the distribution of values and the existence of moments. This property is particularly useful in Finance where for many assets and market indices, the development of models that are able to produce accurate forecasts of future returns or price movements, especially sudden considerable variations, is typically very difficult. 

Moreover, ApEn has been found to be a useful marker of system stability,
with rapid increases foreshadowing significant changes in a time-series \cite{pincus04}. 

Even if we cannot construct a relatively accurate model of the data,
we can still quantify the irregularity of data, and changes thereto,
straightforwardly. Of course, subsequent modeling remains of
interest, although the point is that this task is quite distinct from the
application of effective discriminatory tools. This perspective seems
especially important given the empirical, non-experimental nature
of financial time series.

Representative Applications. Because most financial analyses and
modeling center on price increments or returns, rather than on
prices, we do so below. Given a series of prices {si}, we consider the
%incremental series ui  si1  si; the returns series ri  (si1si)
%1; and the log-ratio series Li  log(si1si). These series are
prominent in evaluating ‘‘random walk’’-type hypotheses. We apply
these series to a variety of assets and indices to illustrate a breadth
of this application mode. We note that in general theoretical and
empirical settings (approximate mean stationarity), ApEn values of
these series are quite similar (e.g., Fig. 1).
Nonetheless, because ApEn can discern shifts in serial characteristics, apart from the consideration of randomness hypotheses,
we expect that application of ApEn to price series {si} directly will
prove useful in clarifying additional changes. Indeed, most empirical applications of ApEn have been to raw time series.

\section{Descriptive Statistics}
\label{desc}

Tables \ref{allprice} and \ref{allreturn} respectively present descriptive statistics For the distribution of closing prices and log-returns. Though the closing price of Bitcoin presents orders of magnitude higher standard deviation and coefficient of variation than that of Gold and the S\&P 500, it is significantly more predictable and presents a lower approximate entropy. Concerning log-returns, the three assets present a roughly similar approximate entropy.

\begin{table}[!htbp]
\begin{center}
\resizebox{\columnwidth}{!}{
\begin{tabular}{|c||c|c|c||c|c|c|c|c|c|}

\cline{2-10}
\multicolumn{1}{c}{} &\multicolumn{3}{|c|}{Daily}&\multicolumn{3}{c|}{Weekly}&\multicolumn{3}{c|}{Monthly}\\
\hline

\cellcolor[gray]{0.8} \textbf{Statistic}&\cellcolor[gray]{0.8} \textbf{Bitcoin} & \cellcolor[gray]{0.8} \textbf{Gold} & \cellcolor[gray]{0.8} \textbf{S\&P 500}&\cellcolor[gray]{0.8} \textbf{Bitcoin} & \cellcolor[gray]{0.8} \textbf{Gold} & \cellcolor[gray]{0.8} \textbf{S\&P 500}&\cellcolor[gray]{0.8} \textbf{Bitcoin} & \cellcolor[gray]{0.8} \textbf{Gold} & \cellcolor[gray]{0.8} \textbf{S\&P 500}\\
\hline
№ of observations&2314&2314&2314&328&328&328&77&77&77\\
\hline
Mean&5149.55&1346.72&2561.64&4915.74&1569.77&2563.51&5861.00&1588.53&2595.77\\
\hline
Standard deviation&5468.10&218.44&467.27&4802.57&518.07&469.74&7111.74&532.30&494.46\\
\hline
coefficient of variation&1.061&0.162&0.182&0.976&0.330&0.183& 1.213&0.335&0.190\\
\hline
Approximate entropy&0.095&0.139&0.137&0.313&0.514&0.315&0.370&0.672&0.417\\
\hline
Kurtosis&6.961&0.256&-0.656& 1.098&0.366& -0.647&1.671&3.184&10.234\\
\hline
\end{tabular}

}
\caption{\label{allprice}Descriptive statistics for closing prices }

\end{center}
\end{table}

\begin{table}[!htbp]
\begin{center}
\resizebox{\columnwidth}{!}{
\begin{tabular}{|c||c|c|c|c|c|c|c|c|c|}

\cline{2-10}
\multicolumn{1}{c}{} &\multicolumn{3}{|c|}{Daily}&\multicolumn{3}{c|}{Weekly}&\multicolumn{3}{c|}{Monthly}\\
\hline

\cellcolor[gray]{0.8} \textbf{Statistic}&\cellcolor[gray]{0.8} \textbf{Bitcoin} & \cellcolor[gray]{0.8} \textbf{Gold} & \cellcolor[gray]{0.8} \textbf{S\&P 500}&\cellcolor[gray]{0.8} \textbf{Bitcoin} & \cellcolor[gray]{0.8} \textbf{Gold} & \cellcolor[gray]{0.8} \textbf{S\&P 500}&\cellcolor[gray]{0.8} \textbf{Bitcoin} & \cellcolor[gray]{0.8} \textbf{Gold} & \cellcolor[gray]{0.8} \textbf{S\&P 500}\\
\hline
№ of observations&2313&2313&2313&327&327&327&76&76&76\\
\hline
Mean&0.00190&0.00016&0.00027&0.01276&0.00099&0.00189&0.05771&0.00879&0.00821\\
\hline
Standard deviation&0.03896&0.00901&0.00968&0.10386&0.05754& 0.02416&0.20962&0.13171&0.04194\\
\hline
coefficient of variation&20.498&	53.0248&	35.426&	8.135&	57.973&	12.73147929&	3.631&	14.971&	5.105\\

\hline
Approximate entropy&1.616&1.610&1.484& 1.084&1.178& 1.138&0.474&0.4972&0.470\\
\hline
Kurtosis&13.023&3.388&20.787& 1.671&3.184& 10.234&-0.243&1.525&1.259\\
\hline
\end{tabular}

}
\caption{\label{allreturn}Descriptive statistics for log-returns }

\end{center}
\end{table}

\begin{figure}[H]
\centering

\begin{subfigure}[b]{\textwidth}
\includegraphics[width=.33\textwidth]{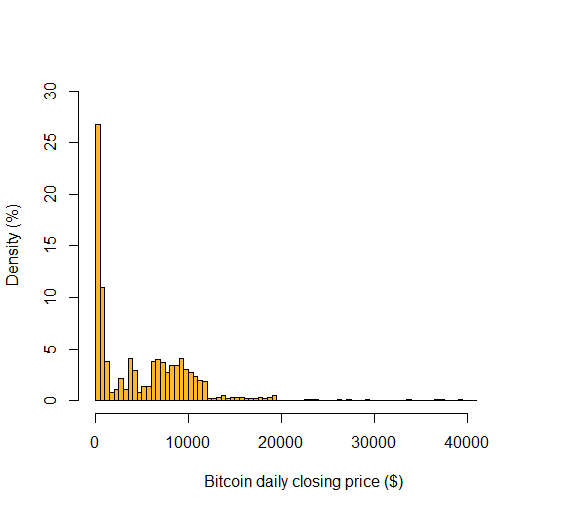}\hfill
\includegraphics[width=.33\textwidth]{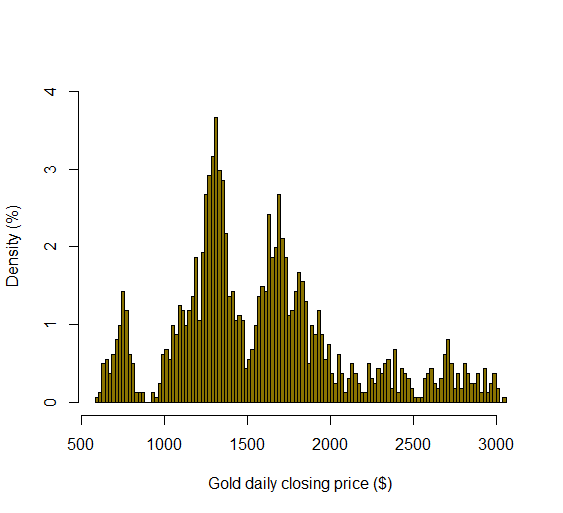}\hfill
\includegraphics[width=.33\textwidth]{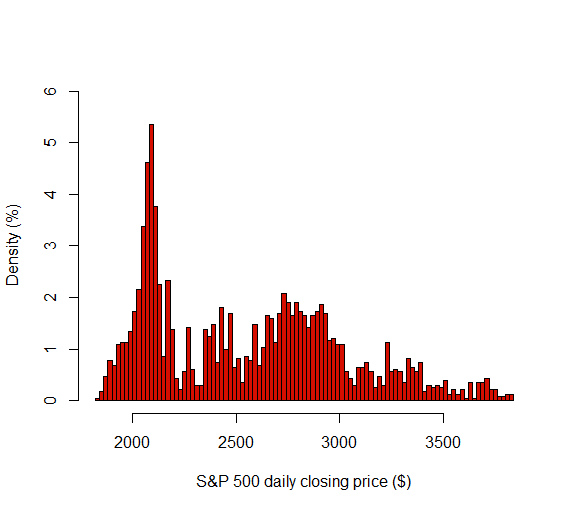}\hfill

\caption{ Histograms of daily closing prices}

\end{subfigure}

\begin{subfigure}[b]{\textwidth}
\includegraphics[width=.33\textwidth]{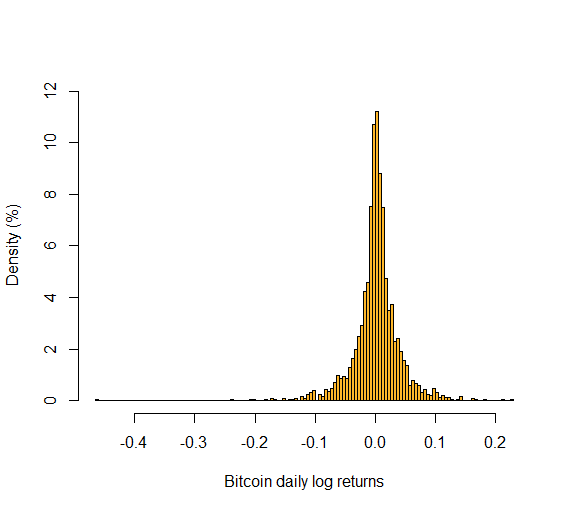}\hfill
\includegraphics[width=.33\textwidth]{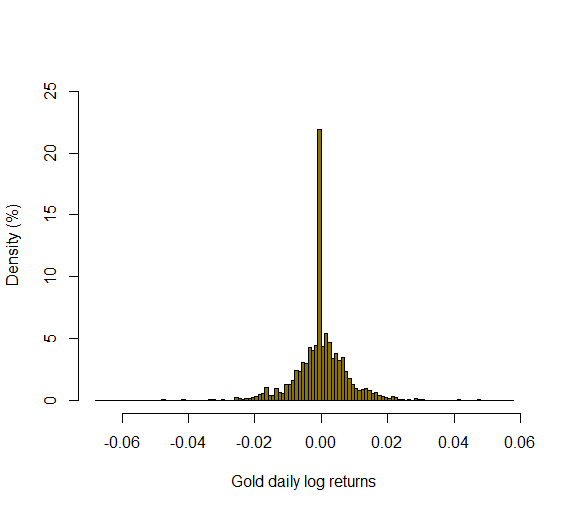}\hfill
\includegraphics[width=.33\textwidth]{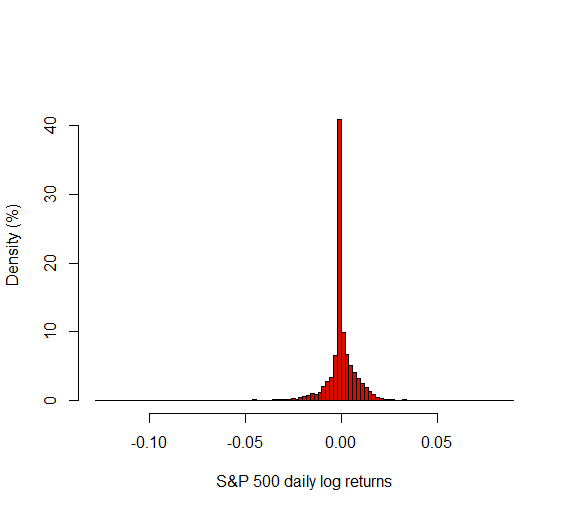}\hfill

\caption{\label{histdailyreturn} Histograms of daily log-returns}

\end{subfigure}

\caption{\label{histdaily} Histograms for daily data}

\end{figure}

In Figure \ref{histdaily} We observe that Bitcoin's closing price presents a long right-tail, when the closing prices of Gold and S\&P 500 are thin-tailed. However, for the three assets, the distributions of log-returns present both left and right long tails. Bitcoin standing out with a multi-modal distribution, when Gold and S\&P 500 are both strongly unimodal with mode at zero.  

\begin{figure}[H]
\centering
\begin{subfigure}[b]{\textwidth}

\includegraphics[width=.33\textwidth]{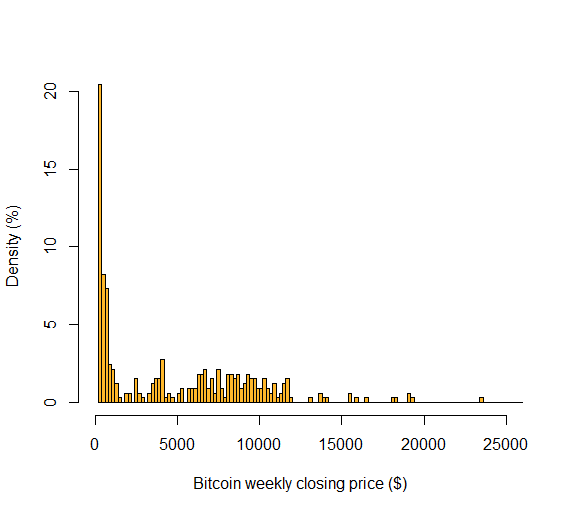}\hfill
\includegraphics[width=.33\textwidth]{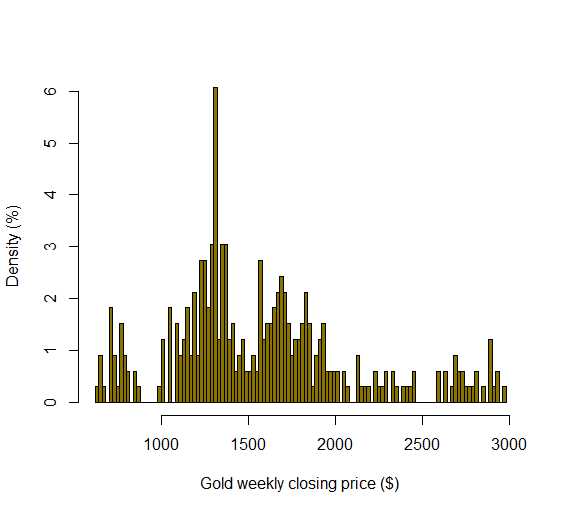}\hfill
\includegraphics[width=.33\textwidth]{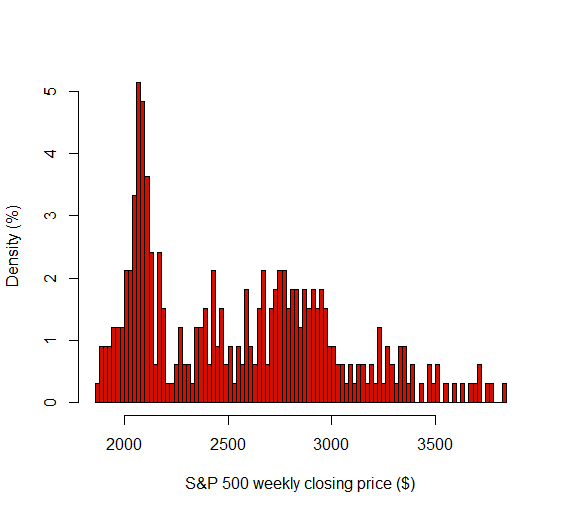}\hfill
\caption{Histograms of weekly closing prices}
\label{histweekly}
\end{subfigure}
\begin{subfigure}[b]{\textwidth}
\includegraphics[width=.33\textwidth]{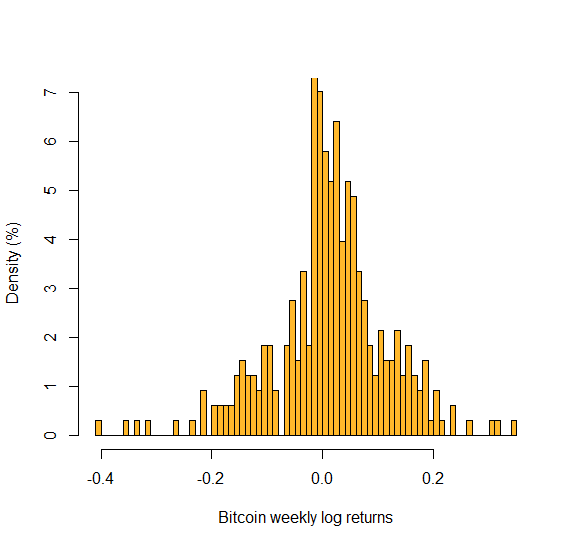}\hfill
\includegraphics[width=.33\textwidth]{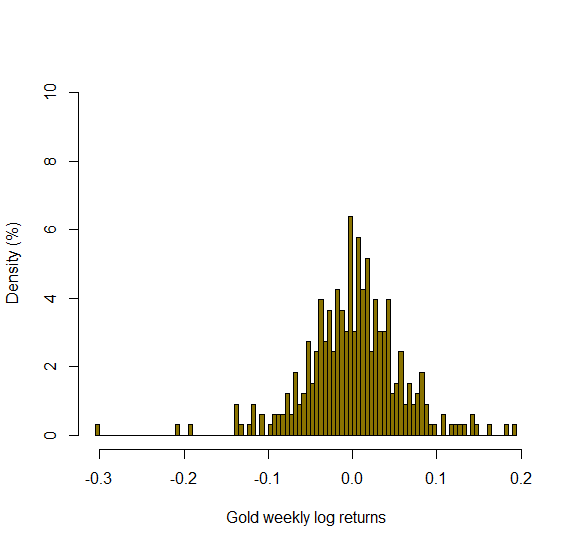}\hfill
\includegraphics[width=.33\textwidth]{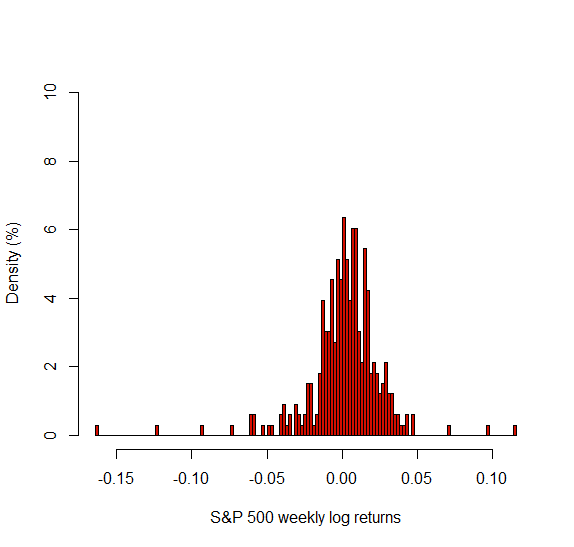}\hfill
\caption{Histograms of weekly log-returns}
\label{histweeklyreturn}
\end{subfigure}
\caption{\label{histweekly} Histograms for weekly data}

\end{figure}

\begin{figure}[H]
\centering
\begin{subfigure}[b]{\textwidth}
\includegraphics[width=.33\textwidth]{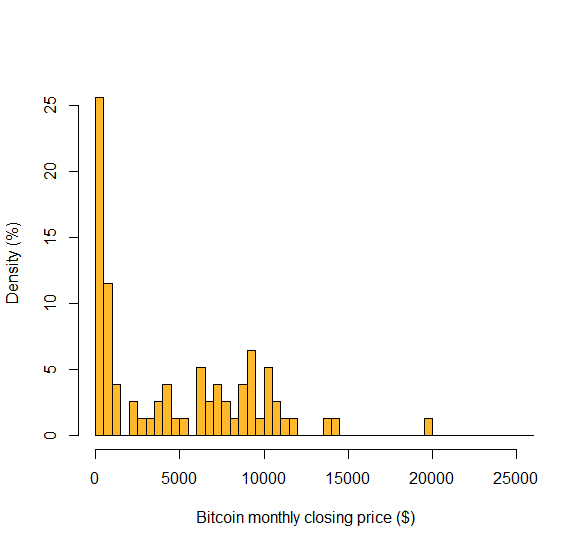}\hfill
\includegraphics[width=.33\textwidth]{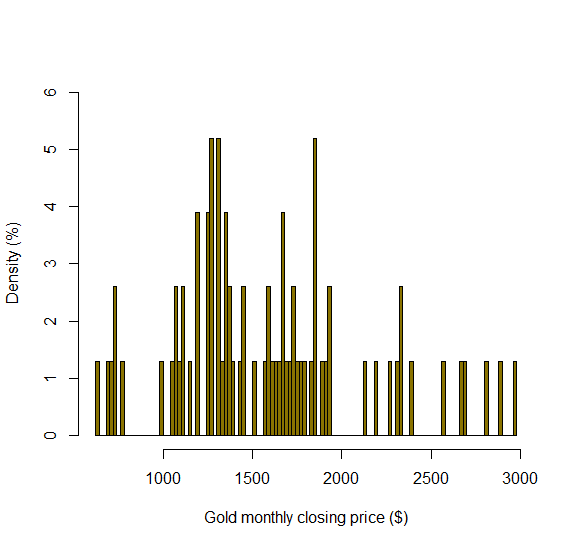}\hfill
\includegraphics[width=.33\textwidth]{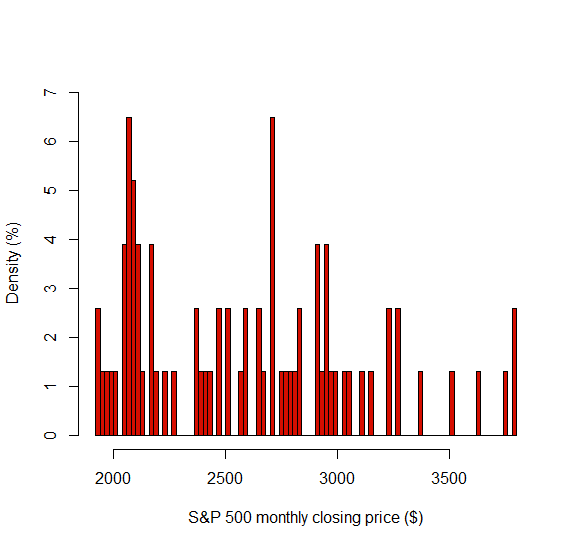}\hfill
\caption{Histograms of monthly closing prices}
\label{histmonthly}
\end{subfigure}
\begin{subfigure}[b]{\textwidth}
\includegraphics[width=.33\textwidth]{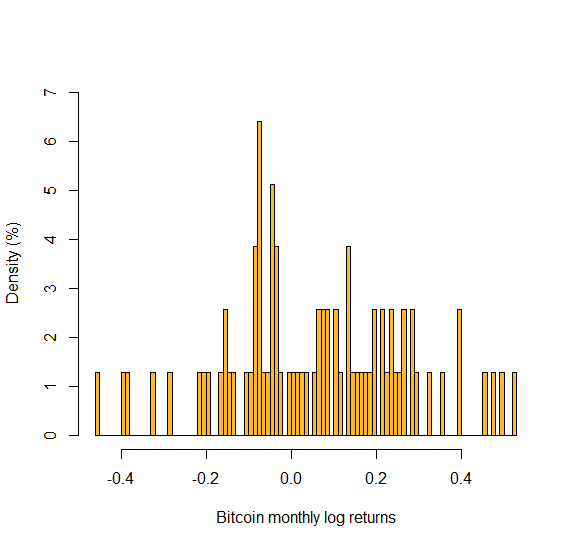}\hfill
\includegraphics[width=.33\textwidth]{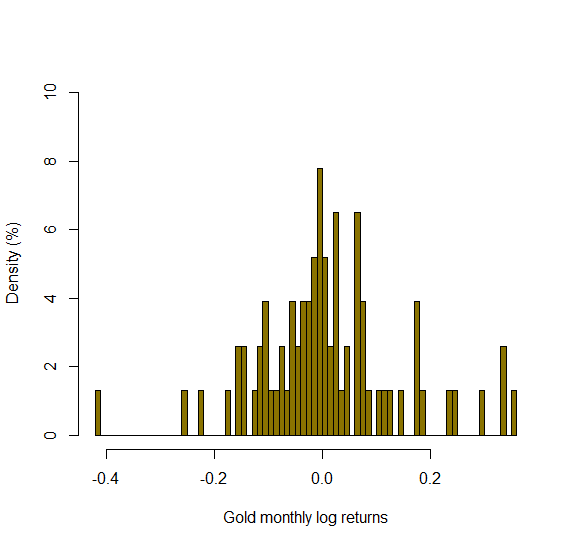}\hfill
\includegraphics[width=.33\textwidth]{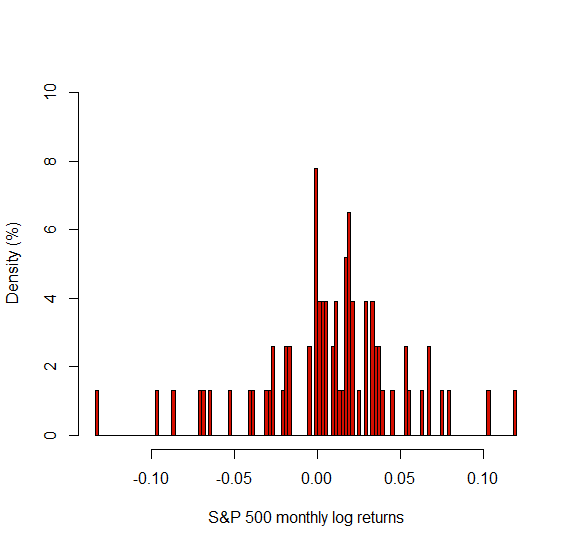}\hfill
\caption{Histograms of monthly log-returns}
\label{histmonthlyreturn}
\end{subfigure}
\caption{\label{histmonthly} Histograms for monthly data}

\end{figure}

Moreover, Histograms for daily log-returns in Figure \ref{histdailyreturn}, weekly log-returns in Figure \ref{histweeklyreturn}, and monthly log-returns in Figure \ref{histmonthlyreturn} show an important dynamic, for different sampling frequencies. Bitcoin's log-returns get progressively more thin-tailed, the longer the time-preference.

This graphical observation is confirmed by the statistics in Table \ref{allreturn}. There is a drastic decrease in the kurtosis of both Bitcoin's closing price and returns, between daily and weekly observations. For the distribution of weekly log-returns, Bitcoin remarkably presents a significantly lower Kurtosis than Gold and S\&P 500, in addition to a lower ApEn and coefficient of variation. Thus, over the complete time-horizon of this study, Bitcoin is less volatile than Gold and S\&P 500 for a weekly time-preference. We will confirm these results using moving coefficients of variation and ApEn in Section \ref{vol}, and further study the tail properties and convergence of moments of the three assets in Section \ref{mef} and Section \ref{maxi}.

\section{Results and discussion}
\label{results}
For different sampling frequencies (daily, weekly, and monthly), corresponding to a progressively longer time-preference, we have studied the tail properties of the closing prices and log-returns of Bitcoin, Gold, and the S\&P 500 index, using two widely used graphical tools from the extreme value analysis; Mean Excess Function plots, to classify the tail behavior of distributions, and Maximum to Sum Ratio plots to study the existence of moments. Moreover, we use the coefficient of variation (standard deviation to mean ratio) and approximate entropy to study the volatility and predictability of closing prices and log-returns.

\subsection{Tail distributions of closing prices and log-returns}
\label{mef}
Consistently with the histograms in Figure \ref{histdaily}, for daily data, the closing price of Bitcoin presents an increasing empirical ME function represented in Figure \label{mefdaily}, past the $\$10,000$ threshold. This indicates that, conditional of having passed that value, the price of Bitcoin is likely to further increase. The daily closing price of Gold and S\&P 500 both exhibit decreasing ME functions consistent with thin-tailed distributions. However, the latter figure shows that the absolute log-returns of all three assets have increasing ME functions consistent with heavy-tailed distributions. Interestingly, the absolute log-returns of S\&P 500 show an ME function that is even more linear than that of Bitcoin, which is consistent with generalized Pareto behavior.

\begin{figure}[H]
\centering
\begin{subfigure}[b]{\textwidth}
\includegraphics[width=\textwidth]{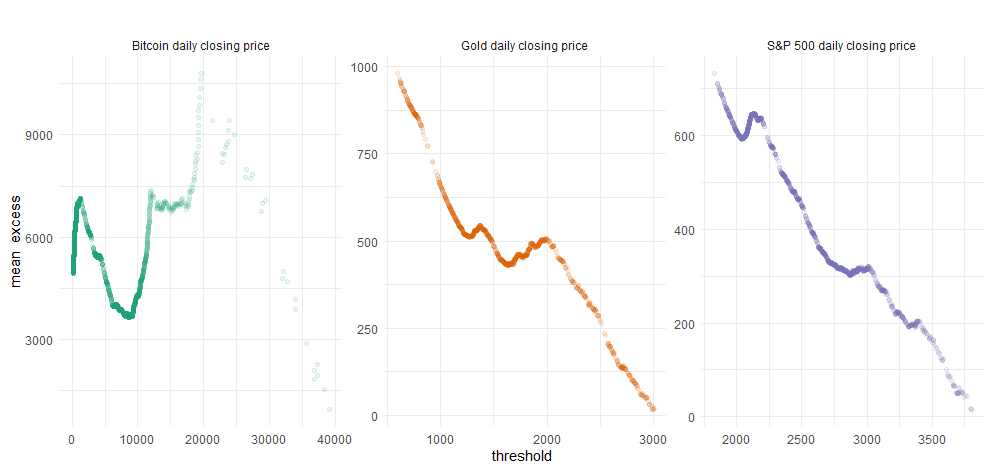}\hfill
\caption{Mean excess plots of daily closing prices}
\end{subfigure}
\begin{subfigure}[b]{\textwidth}
\includegraphics[width=\textwidth]{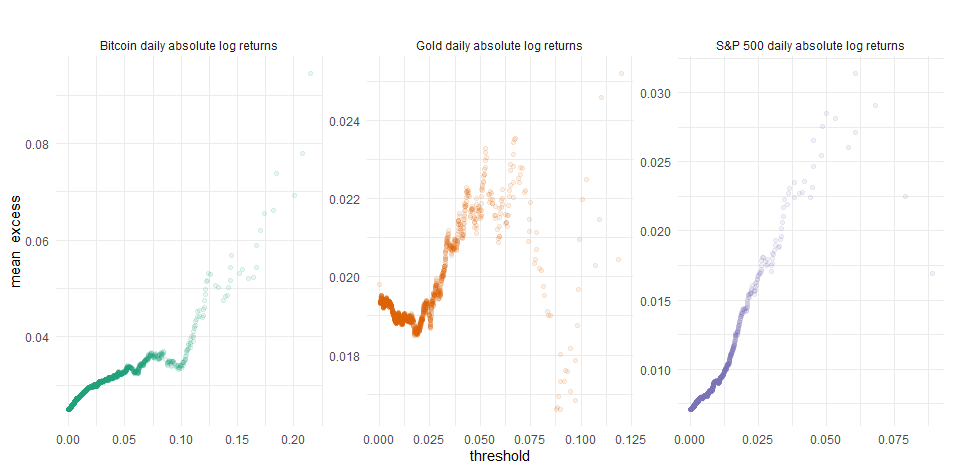}\hfill
\caption{Mean excess plots of daily absolute log-returns}
\end{subfigure}
\caption{Mean excess plots for daily data}
\label{mefdaily}
\end{figure}
\begin{figure}[H]
\centering
\begin{subfigure}[b]{\textwidth}
\includegraphics[width=\textwidth]{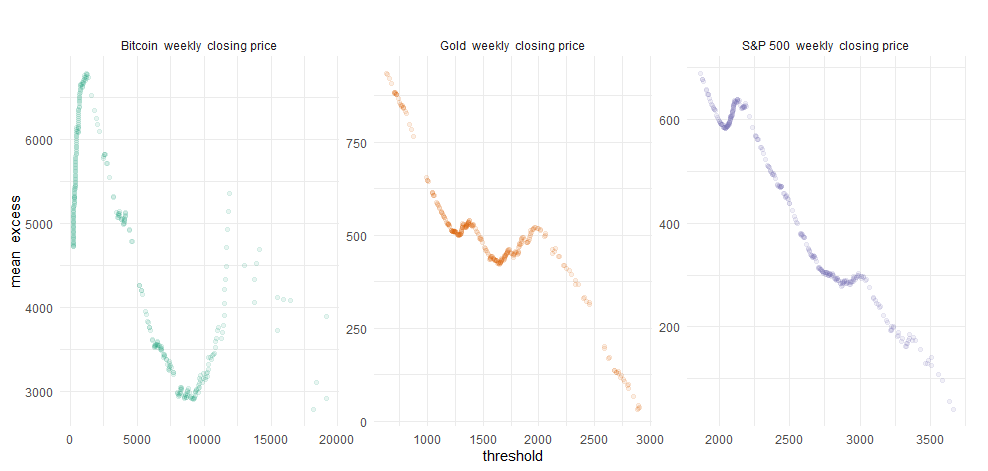}\hfill
\caption{Mean excess plots of weekly closing prices}
\end{subfigure}
\begin{subfigure}[b]{\textwidth}
\includegraphics[width=\textwidth]{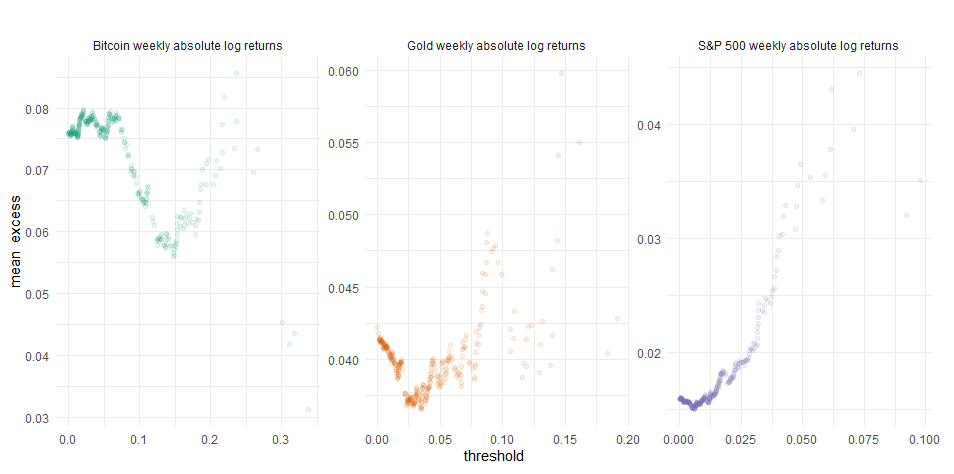}\hfill
\caption{Mean excess plots of weekly absolute log-returns}
\end{subfigure}
\caption{Mean excess plots for weekly data}
\label{mefweekly}
\end{figure}
The ME functions for weekly distributions presented in Figure \ref{mefweekly} generally present the same behavior as daily, with a notable exception; the ME function of Bitcoin's weekly absolute return becomes mainly decreasing for this tsampling frequency, when that of S\&P becomes sharply convex, indicating even more severely heavy-tailed behavior and more disproportionate influence of extrema on moments.

\begin{figure}[H]
\centering
\begin{subfigure}[b]{\textwidth}
\includegraphics[width=\textwidth]{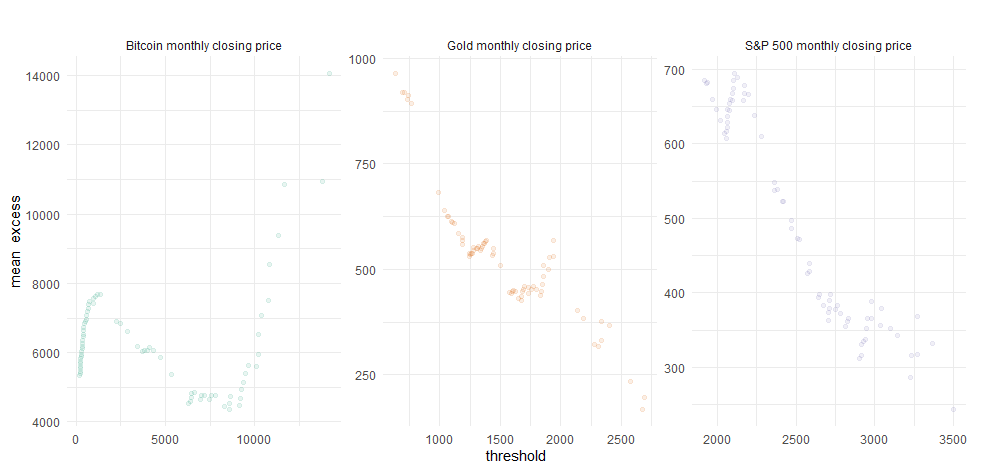}\hfill
\caption{Mean excess plots of monthly closing prices}
\end{subfigure}
\begin{subfigure}[b]{\textwidth}
\includegraphics[width=\textwidth]{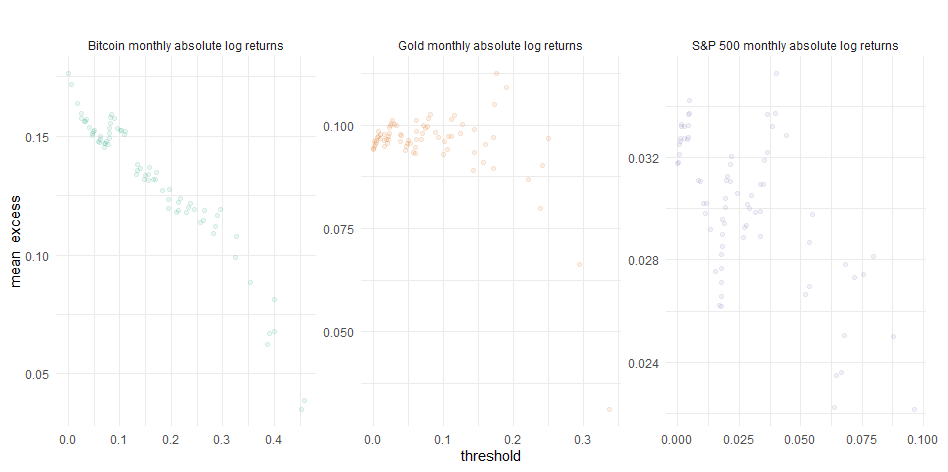}\hfill
\caption{Mean excess plots of monthly absolute log-returns}
\end{subfigure}
\caption{Mean excess plots for monthly data}
\label{mefmonthly}
\end{figure}

Although there only exist 76 observations of monthly closing prices and returns since September 2014, the ME functions of monthly data represented in Figure \ref{mefmonthly} see an accentuation of the previous observations. The distribution of Bitcoin's monthly absolute log-return shows thin-tailed behavior, Gold an ME function that is consistent with an Exponential distribution and S\&P 500, an increasing ME function. 

Thus, for the three sampling frequencues, Bitcoin is the only asset that shows heavy-tailed closing prices. However, the log-returns of Bitcoin get progressively more "well-behaved" and thin-tailed for a longer time-preference, when the opposite effect is observed for S\&P 500's, with returns becoming more extremely heavy-tailed. These findings raise the question of the finiteness of moments, particularly of second moments which are commonly used as indicators of volatility, and of the convergence of their empirical estimators. We address this question in the next section.

\subsection{The convergence of empirical moments}
\label{maxi}
\begin{figure}[H]
\centering
\begin{subfigure}[b]{\textwidth}

\includegraphics[width=\textwidth]{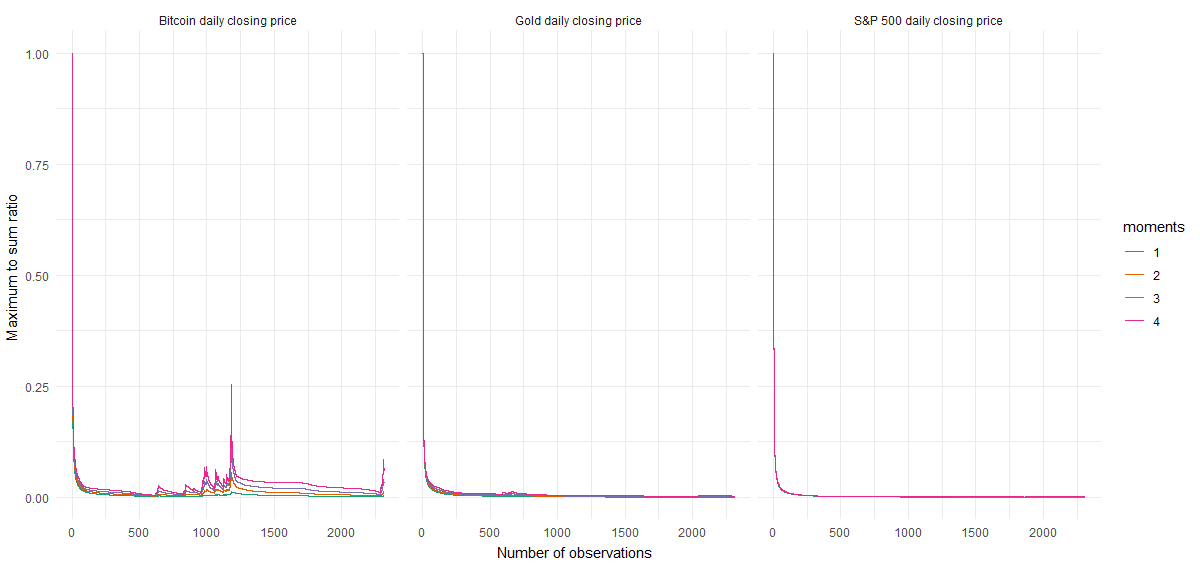}\hfill

\caption{Maximum to sum plots of daily closing prices}
\end{subfigure}
\begin{subfigure}[b]{\textwidth}
\includegraphics[width=\textwidth]{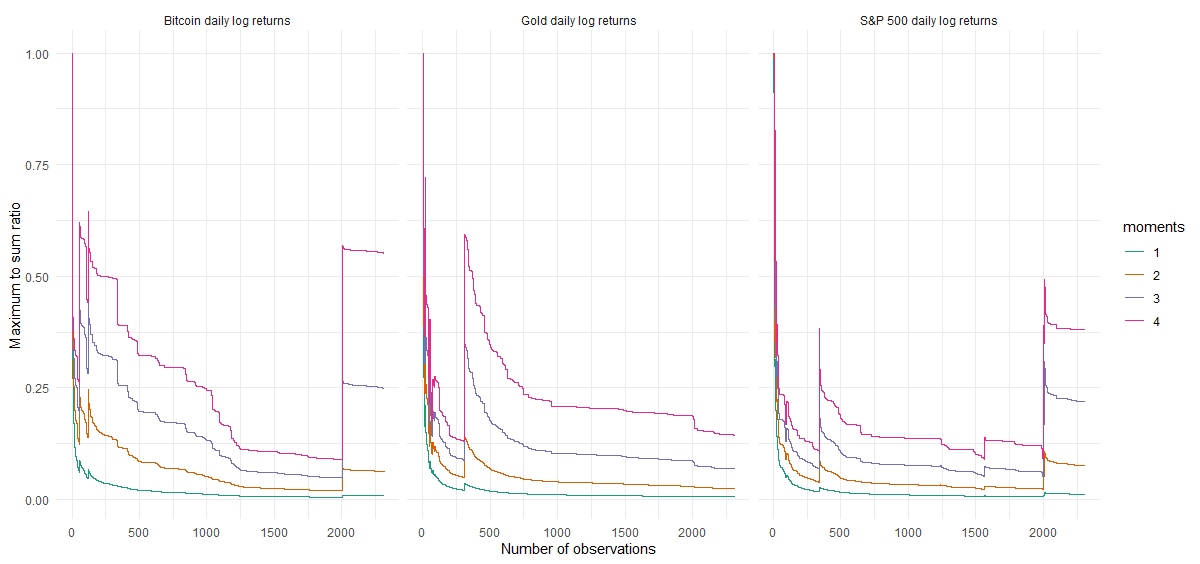}\hfill
\caption{Maximum to sum plots of daily absolute log-returns}
\end{subfigure}
\caption{Maximum to sum plots for daily data }
\label{maxdaily}
\end{figure}

Figure \ref{maxdaily} presents the maximum to sum ratios of daily closing prices and absolute log-returns. As noted in the previous section, the price of Bitcoin shows heavy-tailed behavior and thus slower convergence of moments compared to the other two assets. Moreover, the absolute log-returns of both Bitcoin and S\&P 500 are consistent with generalized Pareto behavior and except for the mean, their higher moments are dominated by extrema and do not converge. Notably, the standard deviations of the log-returns of both assets cannot be reliably estimated from samples.

\begin{figure}[H]
\centering
\begin{subfigure}[b]{\textwidth}
\includegraphics[width=\textwidth]{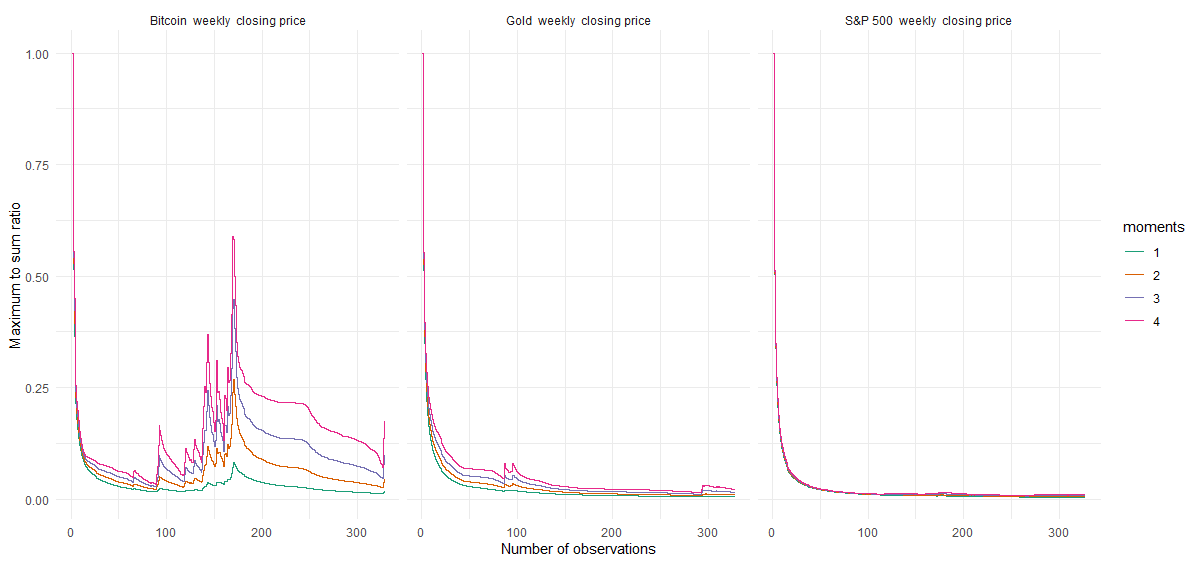}\hfill
\caption{Maximum to sum plots of weekly closing prices}
\end{subfigure}
\begin{subfigure}[b]{\textwidth}
\includegraphics[width=\textwidth]{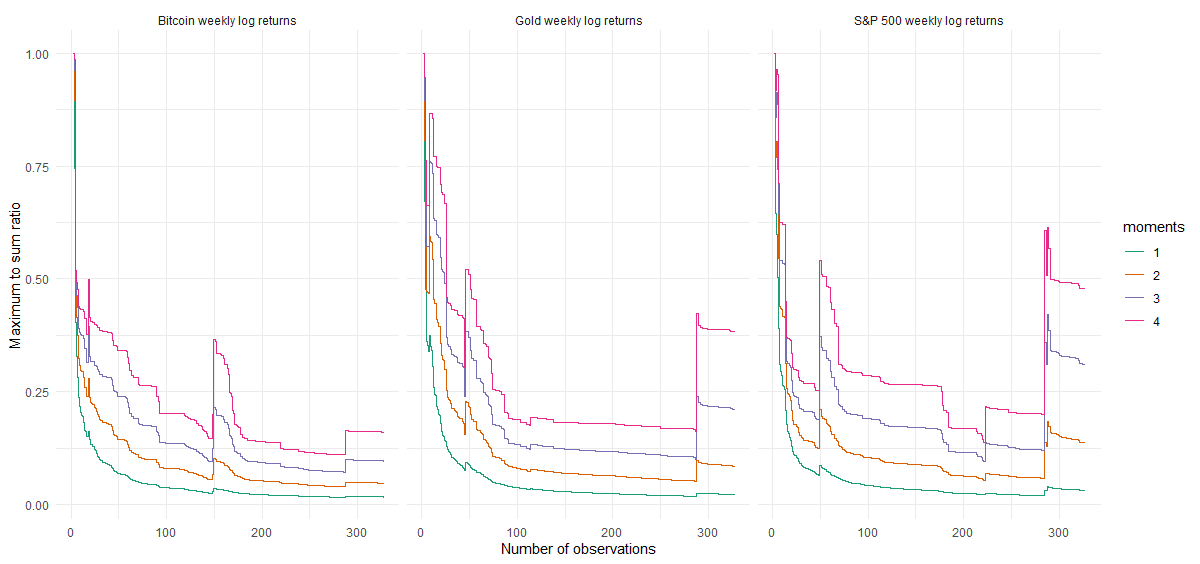}\hfill
\caption{Maximum to sum plots of weekly absolute log-returns}
\end{subfigure}
\caption{Maximum to sum plots for weekly data }
\label{maxweekly}
\end{figure}

\begin{figure}[H]
\centering
\begin{subfigure}[b]{\textwidth}
\includegraphics[width=\textwidth]{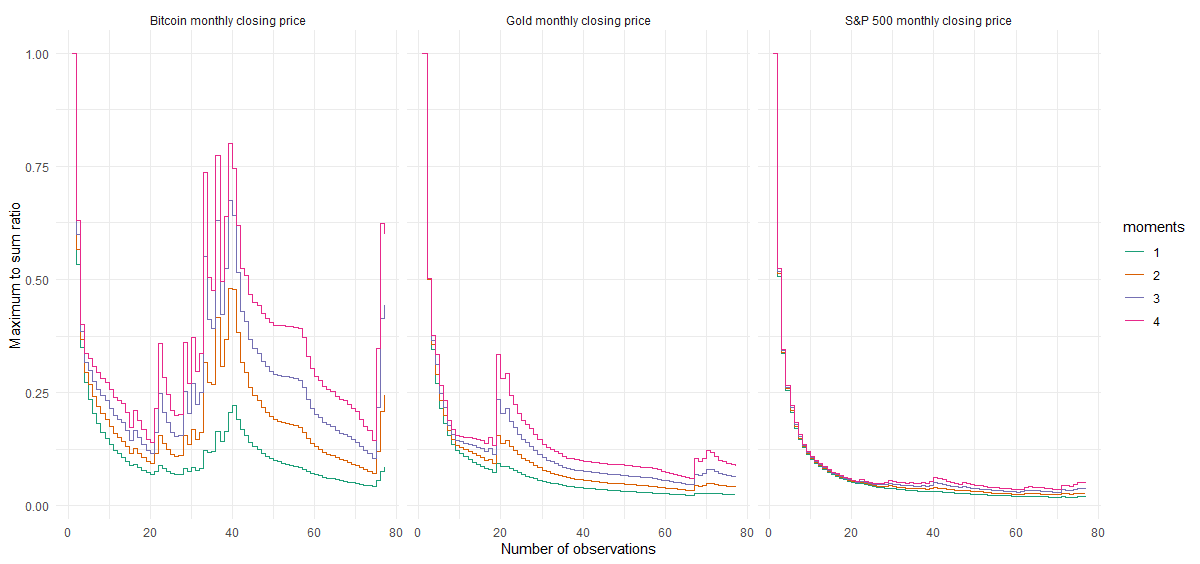}\hfill
\caption{Maximum to sum plots of monthly closing prices}
\end{subfigure}
\begin{subfigure}[b]{\textwidth}
\includegraphics[width=\textwidth]{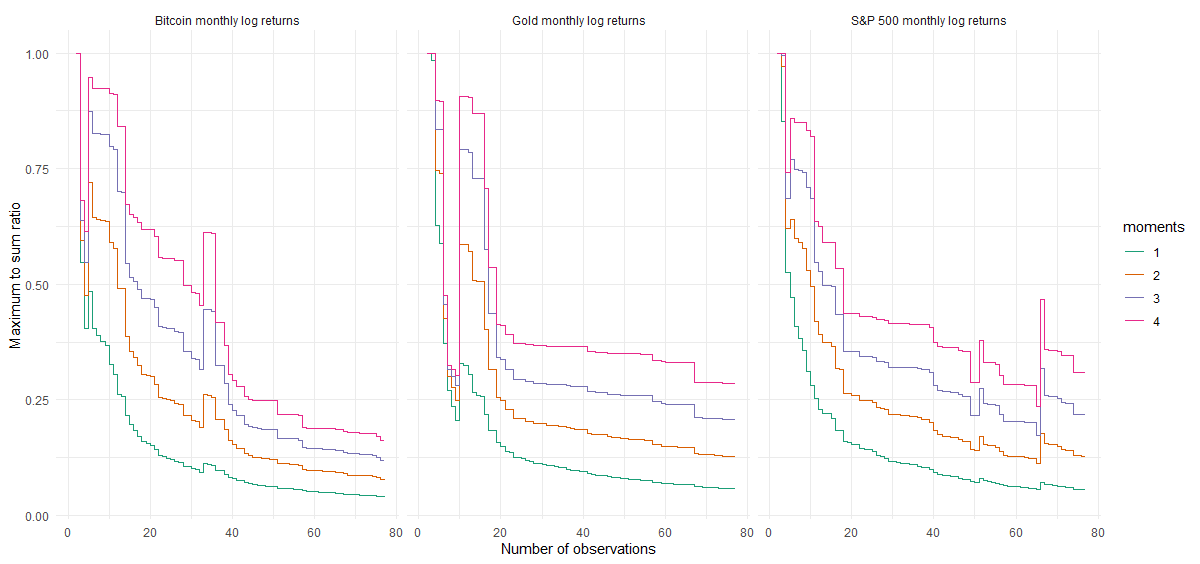}\hfill
\caption{\label{maxmonthlyreturn} Maximum to sum plots of monthly absolute log-returns}
\end{subfigure}
\caption{Maximum to sum plots for monthly data }
\label{maxmonthly}
\end{figure}

Maximum to sum ratios for weekly data in Figure \ref{maxweekly} and monthly data in Figure \ref{maxmonthly} also confirm and precise our observations based on ME functions. A lower sampling frequency make the log-returns of Bitcoin more thin-tailed and empirical moments convergent to their theoretical value. The variance of absolute log-returns of Bitcoin converges even faster than Gold's, for weekly and monthly data, and exhibits a behavior consistent with a log-normal random variable with finite variance for these two sampling frequencies. However, the absolute log-returns of Gold and S\&P 500 adopt a progressively more Paretian behavior for longer time-preferences, and empirical second moments are non-convergent for weekly and monthly data.  
These results inform and nuance our analysis of volatility, in the next section. Non-convergent second moments for the returns of Gold and S\&P 500 mean that the any empirical computation of standard deviation from samples would be dominated by maxima and should be taken as an under-estimate of the real standard deviation of the underlying random variable.

\subsection{Volatility}
\label{vol}
\begin{figure}[H]
\centering
\begin{subfigure}[b]{.32\textwidth}
\includegraphics[width=\textwidth]{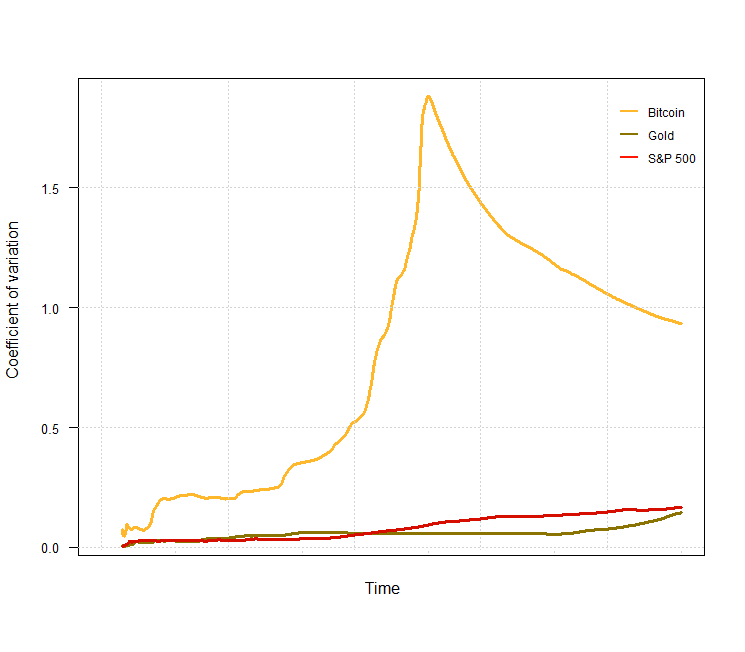}\hfill
\caption{100-day moving coefficient of variation of daily closing prices}
\end{subfigure}
 \hfill
\begin{subfigure}[b]{.32\textwidth}
\includegraphics[width=\textwidth]{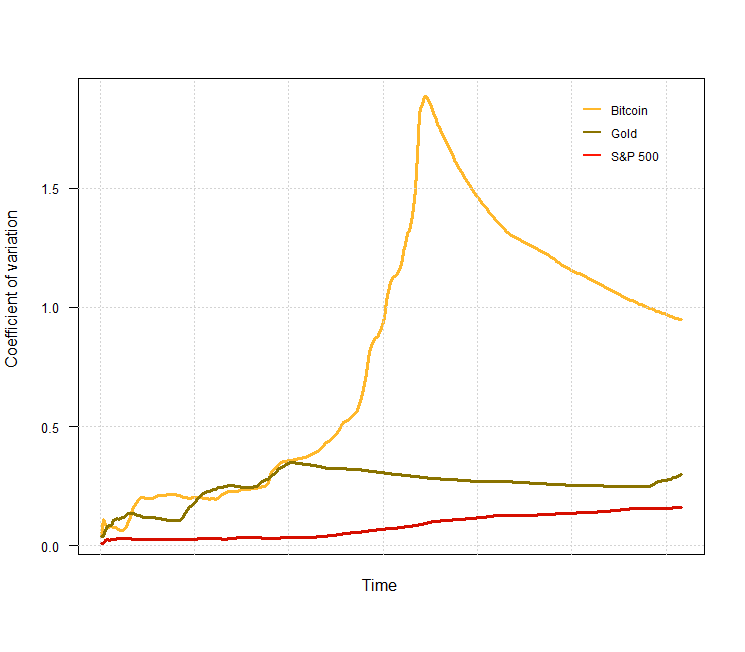}\hfill
\caption{20-week moving coefficient of variation of weekly closing prices}
\end{subfigure}
 \hfill
\begin{subfigure}[b]{.32\textwidth}
\includegraphics[width=\textwidth]{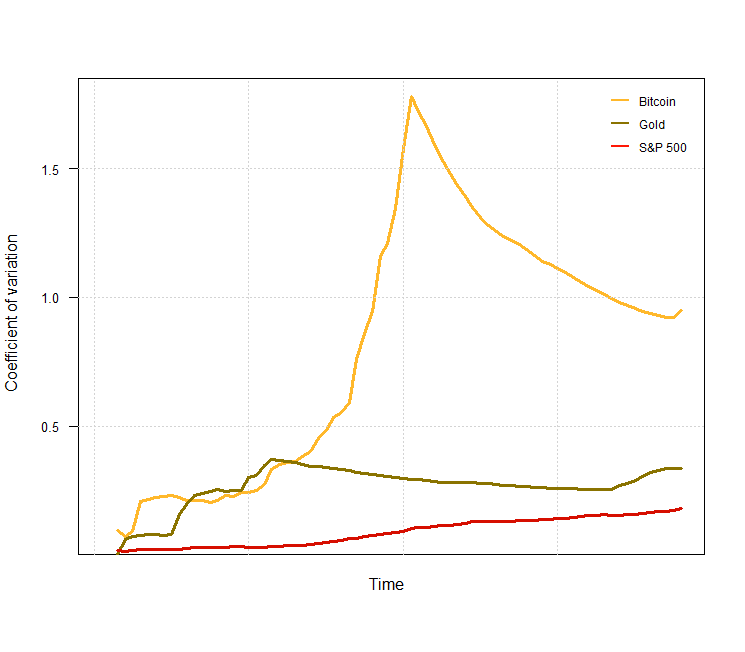}\hfill
\caption{3-month moving coefficient of variation of monthly closing prices}
\end{subfigure}
\caption{Moving coefficients of variation of closing prices. For lower sampling frequencies, sample standard deviation gets progressively higher for Gold, when it remains relatively stable for Bitcoin and S\&P 500.}
\label{sd}
\end{figure}

\begin{figure}[H]
\centering
\begin{subfigure}[b]{.32\textwidth}
\includegraphics[width=\textwidth]{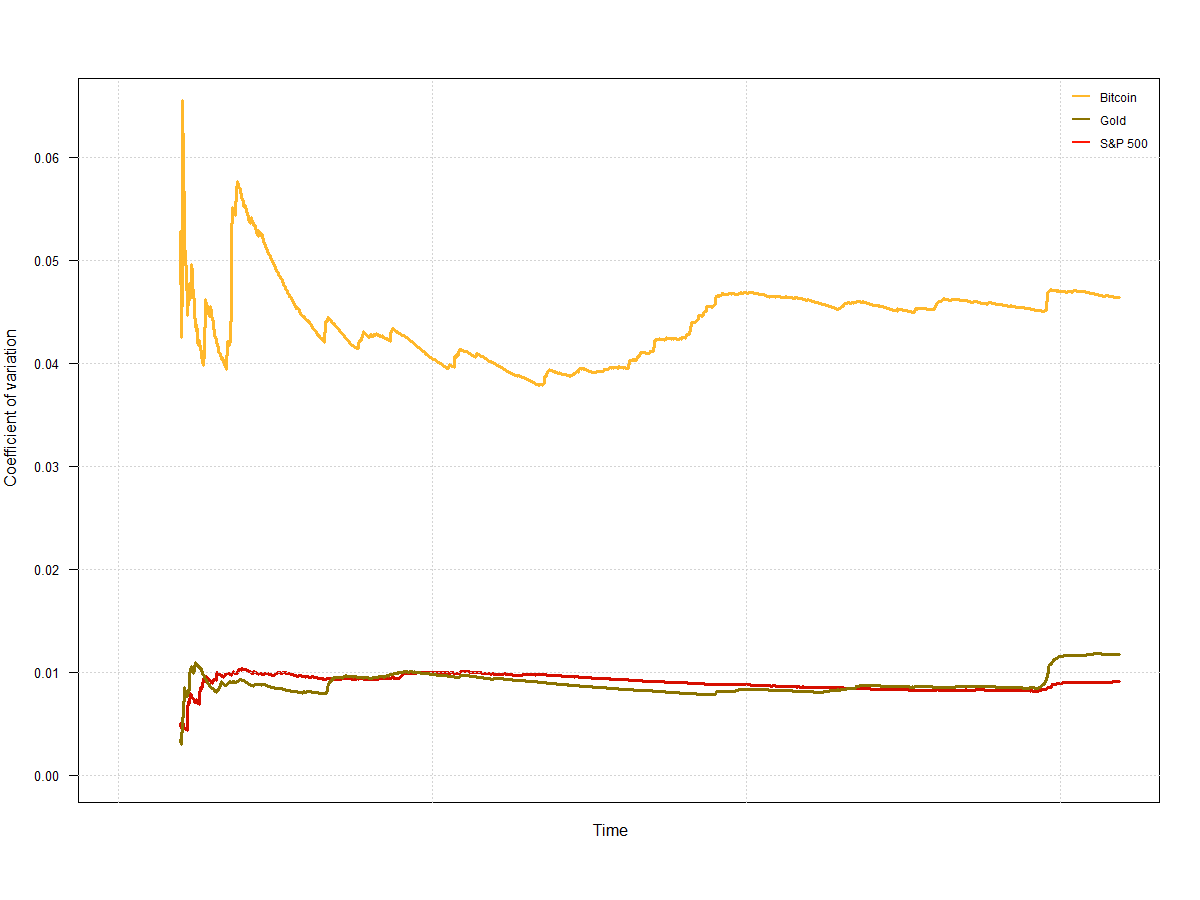}\hfill
\caption{100-day moving coefficient of variation of daily log-returns}
\end{subfigure}
 \hfill
 \begin{subfigure}[b]{.32\textwidth}
\includegraphics[width=\textwidth]{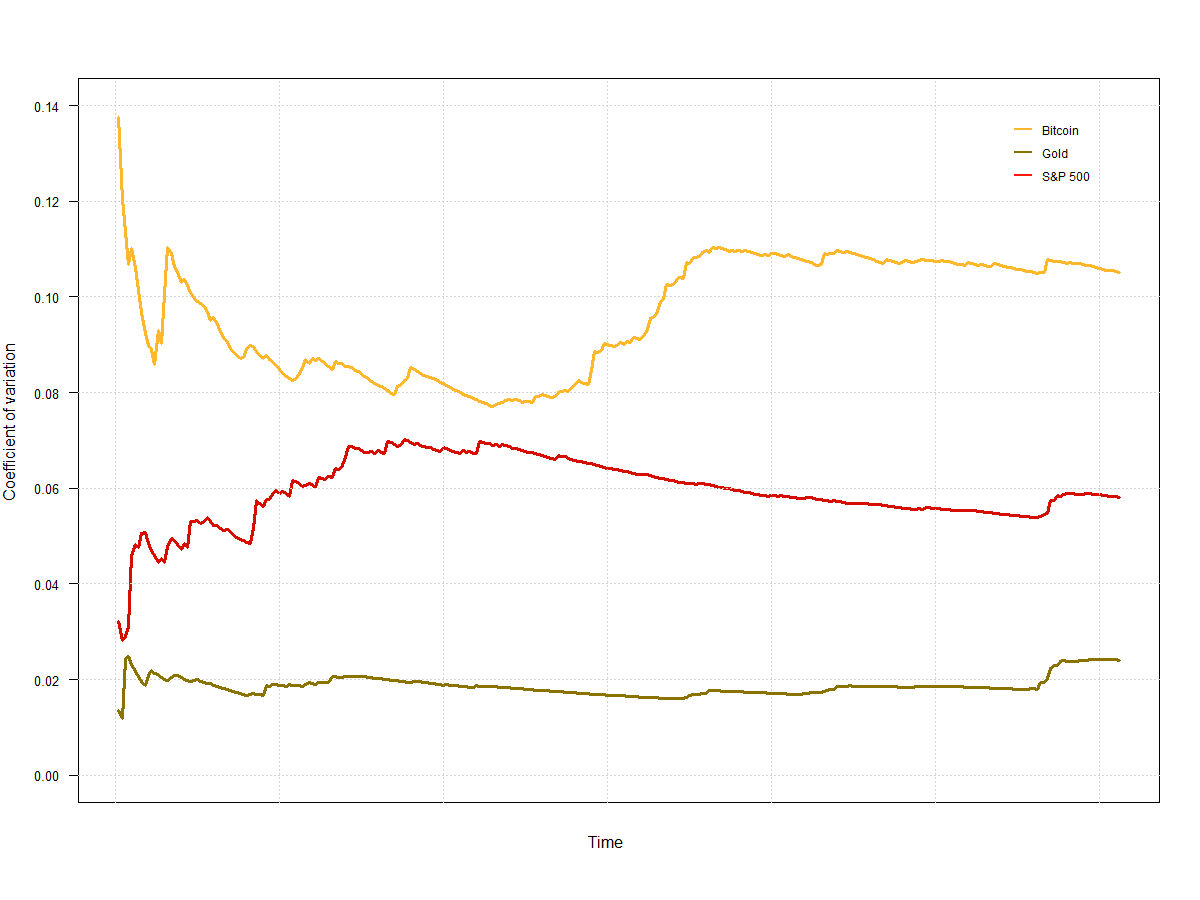}\hfill
\caption{20-week moving coefficient of variation of weekly log-returns}
\end{subfigure}
 \hfill
  \begin{subfigure}[b]{.32\textwidth}
\includegraphics[width=\textwidth]{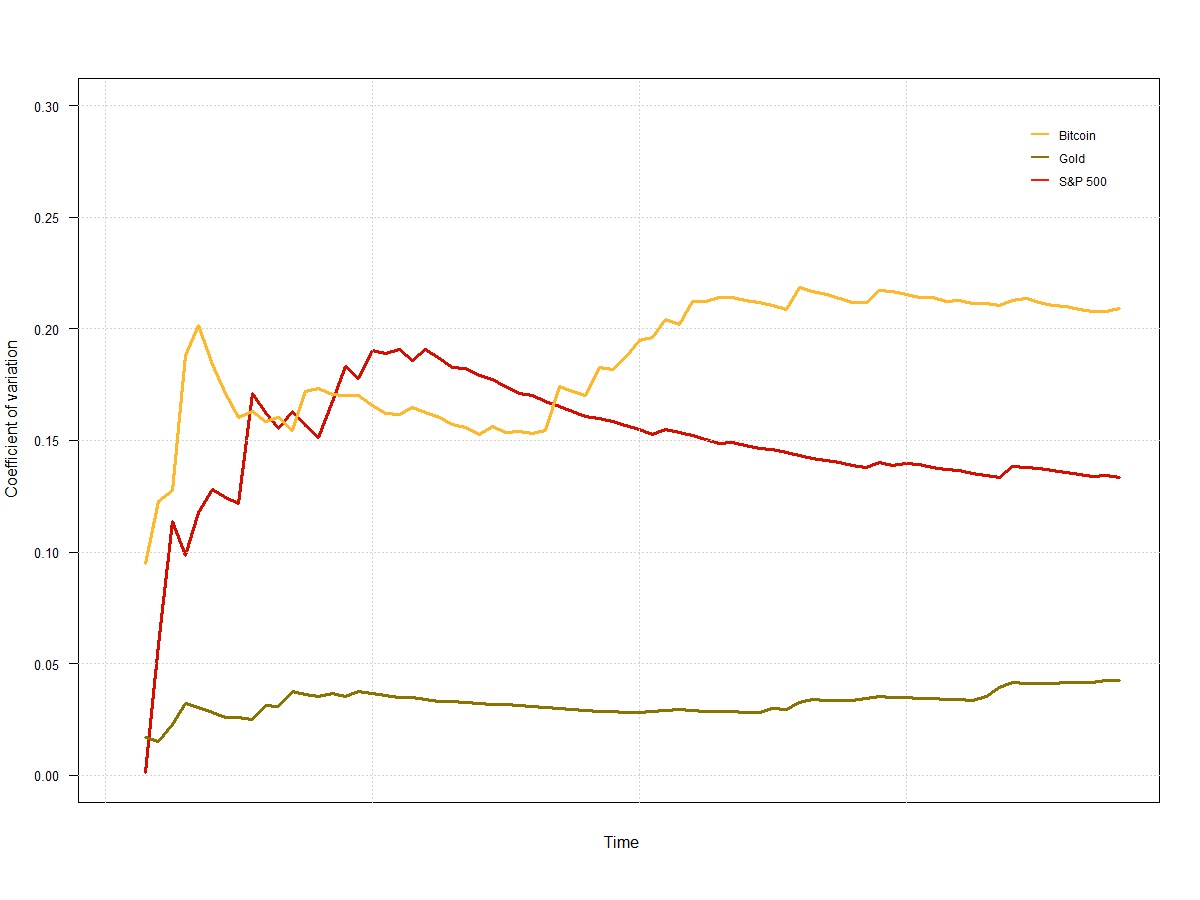}\hfill
\caption{3-month moving coefficient of variation of monthly log-returns}
\end{subfigure}
\caption{Moving coefficient of variation of log-returns. For lower sampling frequencies, sample standard deviation gets progressively higher for S\&P 500, even exceeding Bitcoin's for monthly log-returns, when it remains stable for Bitcoin and Gold. }
\label{sdreturn}
\end{figure}

We have computed 100-day, 20-week, and 3-month rolling standard deviations of the closing prices and log-returns for the three assets. Results are respectively presented in Figure \ref{sd} and Figure \ref{sdreturn}. Unsurprisingly, the closing price of Bitcoin shows the highest moving standard deviation among the three assets in Figure \ref{sd}. However, we observe a significant increase in the moving standard deviation of Gold for lower sampling frequencies when Bitcoin's and S\&P 500's remain relatively stable. 
The same increase is observed concerning the standard deviations of log-returns, with S\&P 500 taking the stead of Gold, while Bitcoin and Gold maintain stable standard deviation for lower sampling frequencies.  Remarkably, the monthly log-returns of S\&P 500 show an even higher rolling standard deviation than Bitcoin's, for several months. Furthermore, this high volatility of monthly S\&P 500 log-returns is likely under-estimate, due to the slower convergence of this random variables empirical second moments (relative to Bitcoin's and Gold), discussed in Section \ref{maxi} and summarized in Figure \ref{maxmonthlyreturn}.  

\begin{figure}[H]
\centering
\begin{subfigure}[b]{\textwidth}
\includegraphics[width=\textwidth]{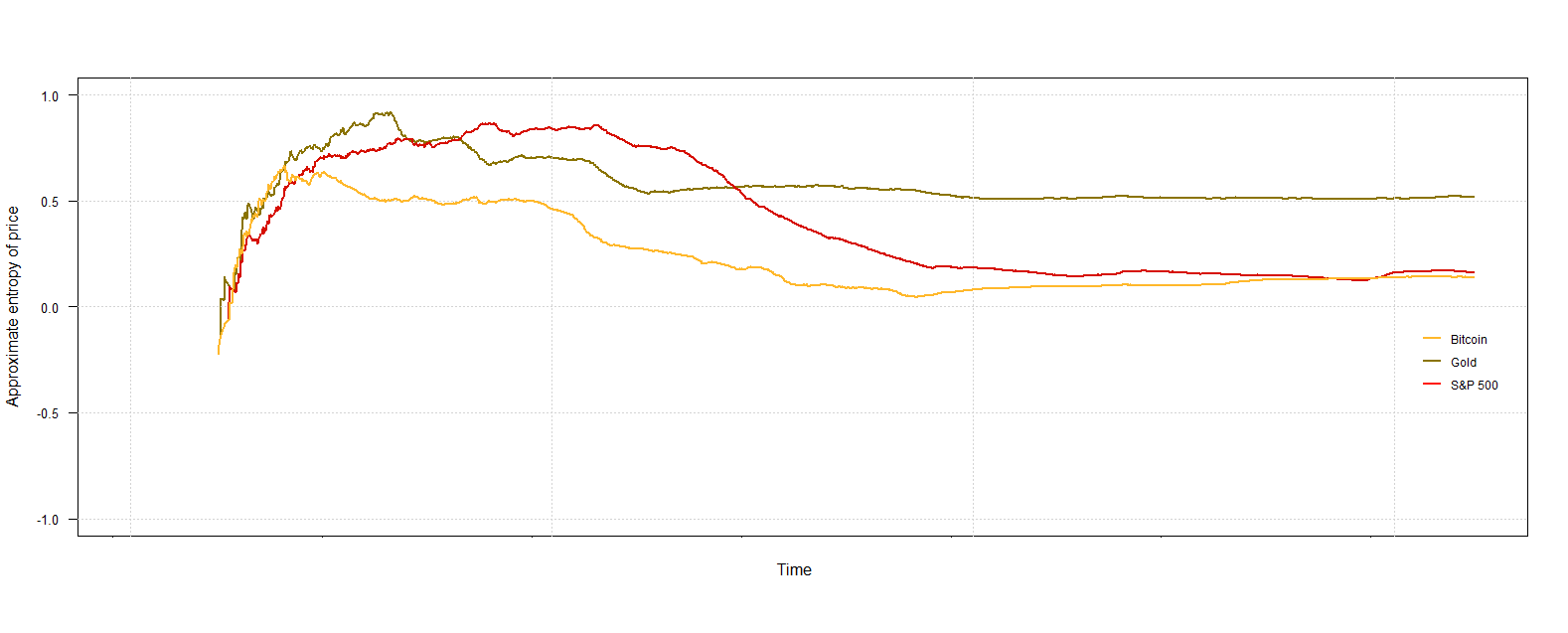}\hfill
\caption{100-day moving Approximate Entropy of daily closing prices}
\end{subfigure}

\begin{subfigure}[b]{\textwidth}
\includegraphics[width=\textwidth]{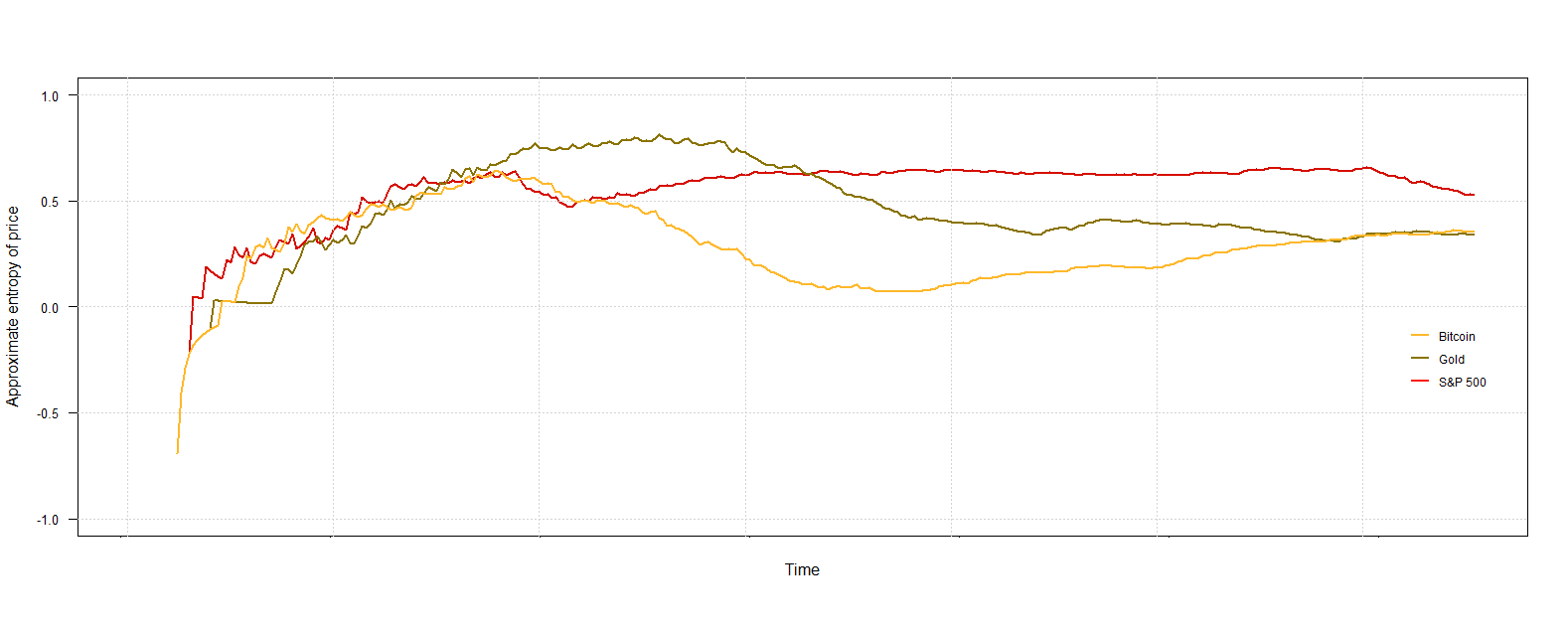}
\caption{20-week moving Approximate Entropy of weekly closing prices}
\end{subfigure}

\hfill
\begin{subfigure}[b]{\textwidth}
\includegraphics[width=\textwidth]{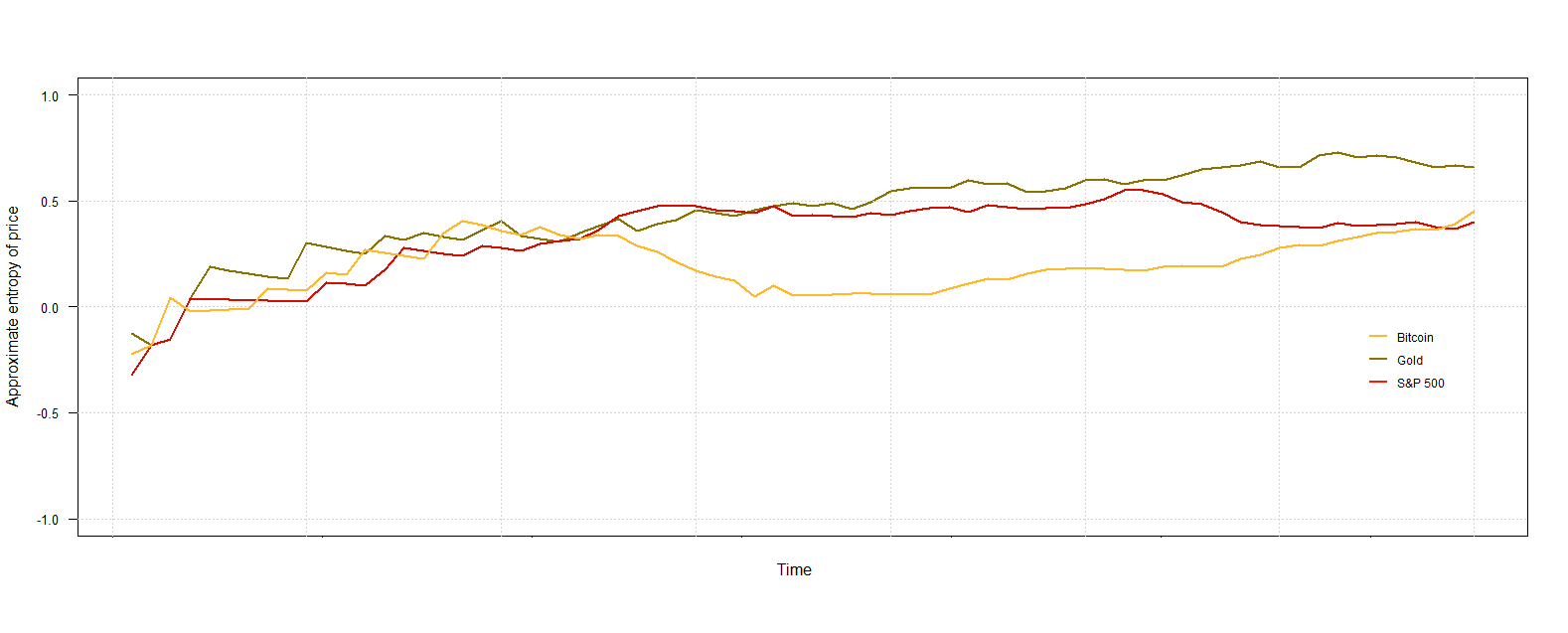}
\caption{3-month moving Approximate Entropy of monthly closing prices}
\end{subfigure}

\caption{Moving Approximate Entropy of closing prices}

\label{apen}
\end{figure}

\begin{figure}[H]
\centering
\begin{subfigure}[b]{\textwidth}
\includegraphics[width=\textwidth]{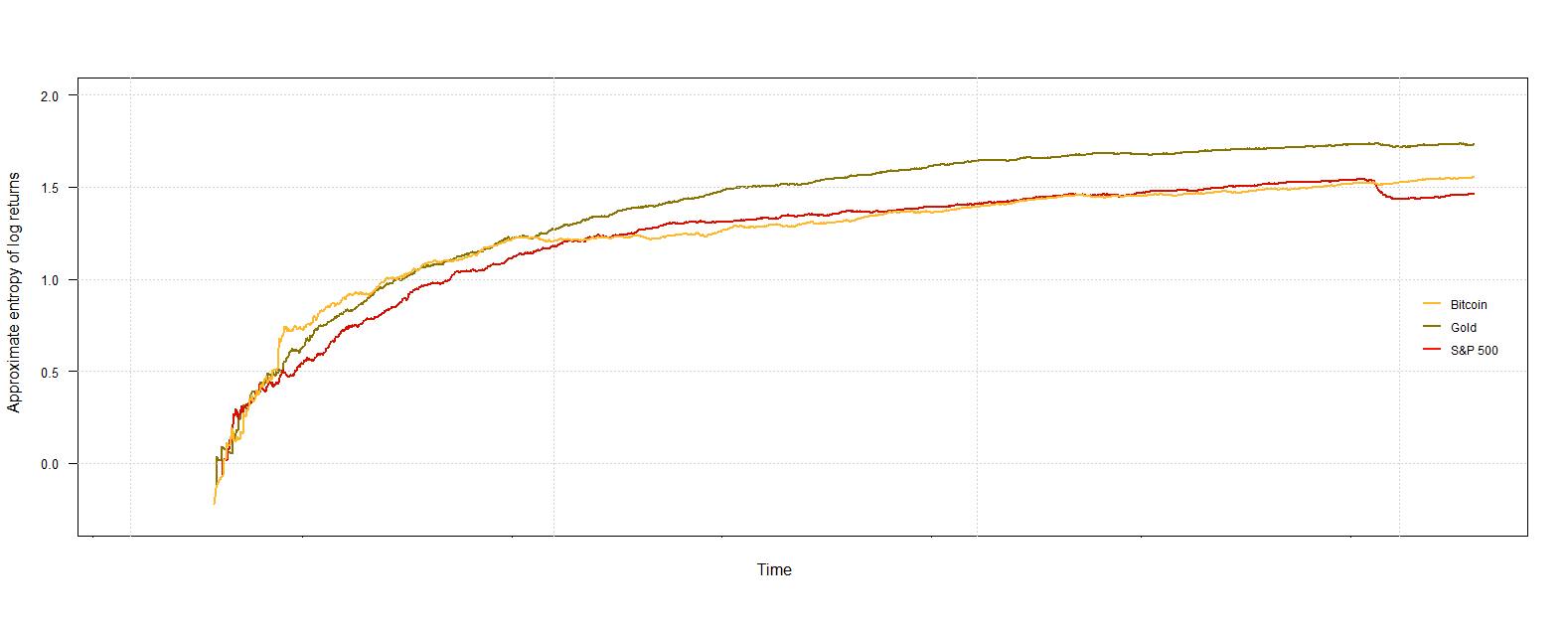}\hfill
\caption{100-day moving Approximate Entropy of daily log-returns}
\end{subfigure}
\begin{subfigure}[b]{\textwidth}
\includegraphics[width=\textwidth]{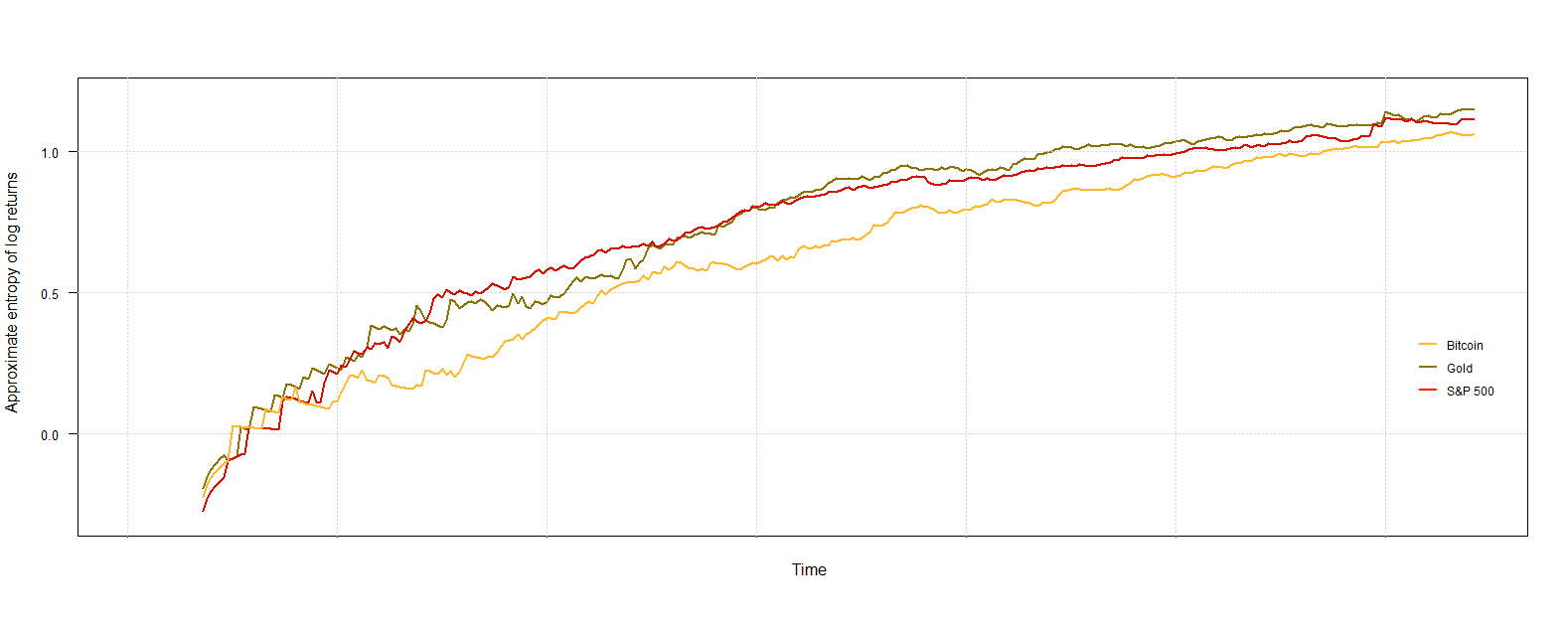}
\caption{20-week moving Approximate Entropy of weekly log-returns}
\end{subfigure}
\hfill
\begin{subfigure}[b]{\textwidth}
\includegraphics[width=\textwidth]{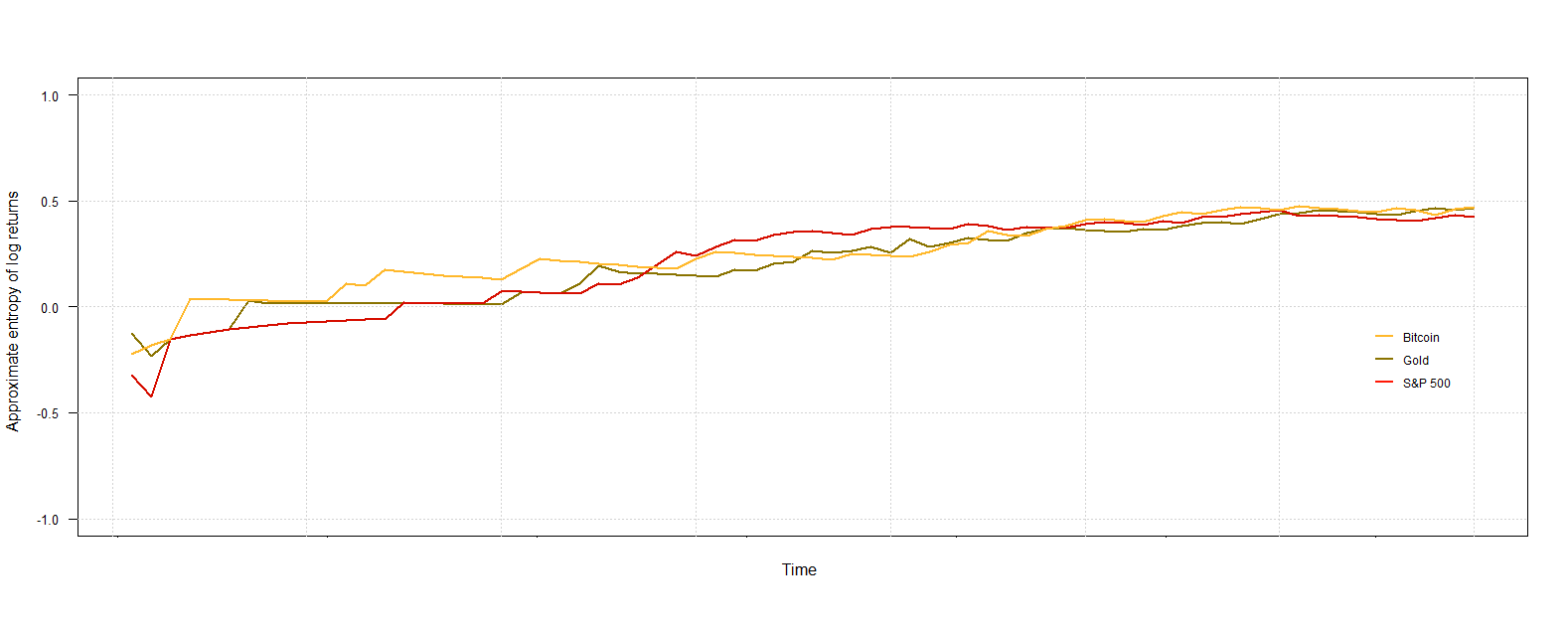}
\caption{3-month moving Approximate Entropy of monthly log-returns}
\end{subfigure}
\caption{Moving Approximate Entropy of log-returns}
\label{apenreturn}
\end{figure}

Given the slow or non-convergence of empirical second moments, as well as the inherent limitations of standard deviation, discussed in Section \ref{toy}, we have complemented our analysis of the three assets'  volatility by computing 100-day, 20-week, and 3-month rolling Approximate Entropies of their closing prices and log-returns. Results are respectively presented in Figure \ref{apen} and Figure \ref{apenreturn}.
Our main finding is that Bitcoin's both the closing price shows a significantly lower moving ApEn than Gold and S\&P 500 for all sampling frequencies. Moreover, Bitcoin's moving ApEn is stable for different time-preferences, when the moving ApEn of the two other assets is, once again, sensitive to the change in the time-scale. This predictability explains the remarkable success and accuracy of Bitcoin price prediction models based on the Stock to Flow ratio \cite{saifedean}. 
Albeit in a less marked way, the log-returns of Bitcoin also exhibit lower ApEn than those of the two other assets. The highest discrepancy in moving ApEn is observed for weekly log-returns. These findings, along with our study of Bitcoin tail behavior and standard deviation, indicate that Bitcoin overall offers a high level of predictability, along with high yield, which make it attractive to investors and speculators, but its high standard deviation excludes its use a currency. 

\section{Conclusion}
\label{conclusion}
Bitcoin still has a long way to go in order to achieve the price stability of Gold so caution is always warranted, but based on our analyses, investors and traders now have empirical evidence for the best time-preference to analyze a Bitcoin investment and what to expect in terms of the behavior closing price and returns. 
If the daily price volatility of Bitcoin excludes its use as a currency, we have shown that the volatility of both its closing price and log-returns significantly decrease for a weekly and monthly sampling frequencies, the log-returns of Bitcoin presenting even thinner tails and comparable coefficient of deviation to S\&P 500 for weekly data. Using approximate entropy, we have shown that, though the variations in the variation in the price and returns of Bitcoin are wide, they exhibit a higher amount of regularity than those of Gold and the S\&P 500. Thus, in the case of Bitcoin, there is a divergence between the Finance and dictionary definition of volatility (“likely to change suddenly and unexpectedly, especially by getting worse”). Bitcoin's price and log-returns are likely to change predictably, typically for the better. 
 The tail behavior and unpredictability of both the closing price and log-returns of the two traditional assets have been shown to be sensitive to the time-scale. As a result of the disproportionate influence of extrema in these heavy-tailed distributions, standard deviations do not converge for the returns of all three variables and can therefore not be accurately estimated from samples. Sample standard deviations should be taken with caution as the Law of Large Number would require an impractically large number of observations to apply.

\end{document}